# Exceptionally High Nonlinear Optical Response in Two-dimensional Type II Dirac Semimetal Nickel di-Telluride (NiTe$_2$)

*Saswata Goswami, Caique Campos de Oliveira, Bruno Ipaves, Preeti Lata Mahapatra, Varinder Pal, Suman Sarkar, Pedro A. S. Autreto\*, Samit K. Ray\* and Chandra Sekhar Tiwary\**

S. Goswami, P. L. Mahapatra

School of Nano Science and Technology, Indian Institute of Technology, Kharagpur, West Bengal-721302, India

S. Sarkar

Department of Materials Engineering, Indian Institute of Technology Jammu, Jammu, Jammu & Kashmir 181221, India

C. C. Oliveira, B. Ipaves, P. A. S. Autreto

Center for Natural and Human Sciences (CCNH)

Federal University of ABC

Rua Santa Adélia 166, Santo André 09210-170, Brazil.

Pedro.autreto@uafbc.edu.br

S. K. Ray

Department of Physics, Indian Institute of Technology Kharagpur, West Bengal 721302, India

E-mail: physkr@phy.iitkgp.ernet.in

V. Pal, C. S. Tiwary

Department of Metallurgical and Materials Engineering, Indian Institute of Technology Kharagpur, West Bengal 721302, India

E-mail: chandra.tiwary@metal.iitkgp.ac.in

Nickel ditelluride (NiTe$_2$) is a newly identified Type-II Dirac semimetal, showing novel characteristics in electronic transport and optical experiments. In this study, we explored the nonlinear optical properties of two-dimensional NiTe$_2$ using experimental and computational techniques (density functional theory-based approach). Few layered two-dimensional NiTe$_2$ (2D-NiTe$_2$) are synthesized using liquid phase exfoliation (LPE), which is characterized using X-ray diffraction technique, transmission electron, and atomic force microscopy. The nonlinear refractive index ($n_2$) and third-order nonlinear susceptibility ($\chi^{(3)}_{total}$) of the prepared 2D-NiTe$_2$ are determined from the self-induced diffraction pattern generated using different wavelengths ($\lambda$= 405, 532, and 650 nm) in the far field. In addition, the diffraction pattern generated by spatial self-phase modulation (SSPM) is further verified by varying concentration (2D-NiTe$_2$ in the IPA solvent), wavelength (of incoming laser beams), and cuvette width (active path length). The high value of third-order nonlinear susceptibility $\chi^{(3)}_{monolayer}$ (in order of ×10$^{-9}$ e.s.u.) determined using SSPM in the 2D-NiTe$_2$ can be attributed to the laser-induced hole coherence effect. Lastly, utilizing the reverse saturable absorption property of 2D-hBN, asymmetric light propagation is also demonstrated in the 2D-NiTe$_2$/2D-hBN heterostructure.

# 1. Introduction

Transition metal dichalcogenides (TMDCs) have recently been recognized as an extensive class of two-dimensional materials (2D). Based on the wide range of electronic properties, various material types classified as semiconductors, semimetals, superconductors, and topological phases are found in the TMDCs.[1] Topological semimetals (TSMs) have recently attracted significant interest in the fields of condensed matter physics, due to their distinct topological properties.[2]

Typical topological semimetals (TSMs), such as the Dirac semimetals, the Weyl semimetals, and the nodal-line semimetals, exhibit conical dispersion and are categorized as type-I TSMs. Graphene is a prime example of a type-I Dirac semimetal since it has valence and conduction bands that meet at precise positions in the first Brillouin zone. Furthermore, it exhibits linear dispersion across the entire momentum space.[3] Type-II TSMs refer to a distinct form of quantum matter.[4] Furthermore, it has been shown that TMDCs possess type-II Dirac fermions that defy the principles of Lorentz symmetry.[5] These fermions exhibit a tilted Dirac cone in momentum space, which is related to the point of intersection.[6] $NiTe_2$ is a type-II Dirac semimetal, that has been developed recently. Unlike other semimetals, $NiTe_2$ exhibits a Dirac cone positioned near the Fermi level.[7] In the case of $NiTe_2$, the Dirac cone is tilted due to broken Lorentz symmetry, leading to the emergence of a Dirac point solely at the junction of the electron and hole pocket.[4a] Similar to $NiTe_2$, there are other materials like $PtSe_2$, $PtTe_2$, and $PdTe_2$ in which electronic structure calculations reveal a tilted band crossing, suggesting type II Dirac fermions.[8] In comparison to other compounds within the series mentioned above, the Dirac node in $NiTe_2$ is much closer to the Fermi level. For this reason, more noticeable indications of topological carriers are found in this material. Nevertheless, in the case of $PtSe_2$, $PtTe_2$, and $PdTe_2$, the Dirac-like points are situated much below the Fermi level.[8] The electronic transport in $NiTe_2$ demonstrates non-saturating linear magnetoresistance, indicating topological metallic behavior. The topological signature of $NiTe_2$ has been explored in recent studies as discussed in subsequent section.

As reported by Xu et al., the quantum oscillations in $NiTe_2$ indicate a low effective mass and nontrivial Berry phase for the carriers, which explains the topological behaviour.[7a] In addition, spin and angle-resolved photoemission spectroscopy (SARPES) provided evidence supporting the presence of topological surface states exhibiting chiral spin texture across a broad energy spectrum in $NiTe_2$.[9] $NiTe_2$ provides an excellent platform for investigating unconventional phenomena in its bulk as well as in lower dimensional form.[10] $NiTe_2$ has been reported with multiple applications, such as ultra-fast THz photodetector[11], flexible photodetector[12], and CO-tolerant electrode for HER and OER applications[13]. NiTe$_2$ can also show strong light-matter interaction while used as a Q-switch 2.8 µm laser.[14] However, the application of NiTe$_2$ in photonics is still in its prefatory state. The robust nonlinear characteristics of two-dimensional materials have drawn interest due to their prospective capabilities in applications of various all-optical devices. This strong nonlinear optical effect can produce interesting phenomena, such as the nonlinear Kerr effect. The nonlinear optical characteristics of a specific material can be assessed using three optical processes: Four-wave mixing, Z scan, and Spatial Self-phase Modulation (SSPM). The optical Kerr effect is a nonlinear optical phenomenon of the third order, where the intensity of the incident light modulates its own phase. SSPM has become an important tool for calculating nonlinear optical parameters like nonlinear refractive index ($n_2$) and third-order nonlinear susceptibility ($\chi^3_{total}$).[15] The first SSPM-related discovery was made on 1D nematic crystal, in 1981.[16] Thirty years later, Wu et al. investigated the SSPM effect in graphene, synthesized through chemical exfoliation.[17] Subsequently, there has been significant increasing interest in SSPM spectroscopy for 2D materials, and many groups have reported the SSPM effect in 2D materials such as 2D Te, NbSe$_2$, MoS$_2$, Bi$_2$Te$_3$, MoTe$_2$, MoSe$_2$, SnS, WSe$_2$.[18] Upon comparison, it was observed that the numerical values of the nonlinear optical parameters obtained from other approaches are similar (Four-Wave Mixing, Z Scan, and SSPM). Although Z-Scan and Four-wave mixing have strict requirements for the optical path. Hendry et al.[19] and Zhang et al.[20] independently published results on the nonlinear optical values of graphene using Four-wave mixing and Z scan, their findings revealed, that the value of third-order nonlinear susceptibility ($\chi^3_{monolayer}$) falls within the range of ~10$^{-7}$ e.s.u., which is very identical to the value obtained through SSPM experiment. As discussed earlier, TMDCs are pivotal materials for their tunable bandgap and optoelectronic properties.[21] In the present work, 2D-NiTe$_2$ was synthesised from the bulk NiTe$_2$ through a cost-effective liquid phase exfoliation method (LPE). The nonlinear optical coefficients such as $n_2$ (nonlinear refractive index) and third order nonlinear susceptibility ($\chi^{(3)}_{total}$) is calculated. The progression of the SSPM pattern with time was examined. Additionally, the distortion within the SSPM pattern was scrutinized, and the change in the refractive index was assessed. Ab initio calculations based on Density Functional Theory (DFT) were carried out to investigate the electronic structure and nonlinear optical coefficients of both bulk and 2D-NiTe$_2$. In addition, bulk and 2D-NiTe$_2$ posess wide range of nonlinear optical responses, 2D-NiTe$_2$ dispersed in the IPA solution can be used for the realization of the all-optical devices by applying the SSPM. In this work, the potential application of 2D-NiTe$_2$ as a photonic diode was also explored.

## 2. Synthesis and Characterization of 2D-NiTe$_2$

High-purity Ni (99.9%) and Te (99.9%) were mixed in a 1:2 molar ratio to get the NiTe$_2$ crystals. Bulk NiTe$_2$ was synthesized using the induction melting at 850°C±10°C (held for 10 min.) in an argon atmosphere, followed by furnace cooling. The bulk material was then annealed at 800°C under an argon atmosphere in a sealed quartz tube using a muffle furnace for 96 hours. The NiTe$_2$ crystallizes in the CdI$_2$ type of trigonal structure.[13, 22] NiTe$_2$ is a layered material, each layer is stacked along the c-axis in which each Ni atom is linked with 6 Te atoms in an octahedral manner.[7a, 13] The bulk crystal was grounded into powder using a mortar pestle. This powder was then probe sonicated in the isopropyl alcohol (IPA) solvent for a 4 hr. period using Rivotek SM250PS probe sonicator. After resting the solution for 12 hr., the suspended solution was then subjected to centrifugation at a speed of 2000 revolutions per minute for 30 minutes using a REMI PR-24 centrifuge machine. The 2D-NiTe$_2$ was subsequently retrieved from the suspension, making this method of synthesizing 2D-NiTe$_2$ simple and economical. The XRD analysis was performed using the Bruker D8 advance with Cu-Kα radiation (K$_α$=1.5406 Å). The XRD pattern of NiTe$_2$ was indexed using JCPDS number 98-064-6914. The sharp peaks that can be seen in the XRD pattern (**Figure 1a**) confirm the crystallinity. In addition, both 2D and bulk NiTe$_2$ show similar crystal structure. The lattice dimensions were determined with values a= b = 3.869 Å and c= 5.308 Å, in line with previously existing literature.[13, 23] A high-resolution transmission electron microscopy (FEI, Themis 60-300, and FEI-Ceta 4k × 4K camera) was used for imaging the exfoliated 2D nanostructure of the NiTe$_2$. Dotted lines of different widths were used to show overlapping 2D flakes type nanostructure on top of each other (as shown in Figure 1b). Figure 1c shows the magnified image of the 2D flake. The inset of Figure 1c displays the Fast Fourier Transform (FFT) pattern associated with the (011) plane. The calculated d spacing associated with the (011) plane was found to be 0.2933 nm (as shown in Figure 1d). An atomic force microscope (Agilent Technologies, 5500) was used to assess the thickness and the lateral width of the 2D sample (as shown in Figure 1e). As indicated in Figure 1f, the thickness distribution reveals that the 2D-NiTe$_2$ nanostructures have an approximate thickness ranging from 0-14 nm, with some particles exhibiting significantly greater thickness. Figure 1g shows the height profile of the 2D flakes, which is observed to be in the range of 80 - 240 nm. The IPA solution containing 2D-NiTe$_2$ was then drop casted on a Si wafer for chemical analysis. PHI 5000 VERSA PROBE III ULVAC PHI (Physical Electronics) was used for the XPS analysis. The XPS spectrum ranges from 850-885 eV, and has a sharp peak corresponding to Ni 2p signal. Figure 1h depicts two sharp peaks, these two sharp signals observed at 855 eV and 873.13 eV correspond to Ni $2p_{3/2}$ and Ni $2p_{1/2}$ binding states.[24] The two shoulder signals observed adjacent to the aforementioned peaks can be traced at 860.29 and 878.54 eV, which may be due to the presence of NiO.[25] The formation of this NiO, may be caused by surface oxidation under ambient conditions. In contrast to other Ni-based TMDCs, nickel exhibits +2 oxidation state in the present study, although $NiTe_2$ crystallizes in a layered structure.[26] The chemical analysis shows core level signals of Te ranging from 568.5 to 589.8 eV, and sharp peaks were observed at the binding energy 574.43 eV and 584.91 eV, which corresponds to Te $3d_{5/2}$ and Te $3d_{3/2}$ of the Te in the $NiTe_2$ (as shown in Figure 1i). Two satellite signals of $TeO_2$ were observed at 576.35 eV and 586.89 eV, which arise due to surface oxidation.[13] Additionally, two satellite peaks were also observed at 571.62 eV and 582.13 eV, indicating the existence of elementary Te, following the sonication process, as the Te-Te bond is comparatively easier to break.[24]

As depicted in Figure 1j, UV-visible spectroscopy was employed to evaluate the optical absorbance of 2D-$NiTe_2$. The inset of Figure 1j shows the Tauc plot that was used to estimate the bandgap of the 2D-$NiTe_2$, which was calculated to be 1.23 eV. Tauc plot is described in the expression as,

$$(\alpha h\nu)^{1/n} = A(h\nu - E_g) \qquad (1)$$

α is the optical absorption coefficient, $h\nu$ is the discrete photon energy, $E_g$ is the bandgap of the material. The value of $n$ is $\frac{1}{2}$ and 2 for Direct and Indirect bandgap respectively. Here direct transition is considered hence the value is considered to be $\frac{1}{2}$. A Raman spectrometer (Witech alpha 300), aided with an optical microscope, was used to determine the vibrational modes of the 2D-$NiTe_2$ nanostructure. The Raman spectroscopy of 2D-$NiTe_2$ is provided in Figure 1l. As observed, two prominent vibrational peaks were detected at 86.29 $cm^{-1}$ and 136.92 $cm^{-1}$, which aligned with the $E_g$ (shown in Figure 1k) and $A_{1g}$ (shown in Figure 1m) vibrational modes. The $E_g$ mode suggested atomic vibrations within the plane, with the top and bottom tellurium (Te) atoms exhibiting opposite movements. Additionally, an out-of-plane $A_{1g}$ phonon mode was detected, signifying a stronger interaction between electrons. It is crucial to emphasize that there is a 50.63 $cm^{-1}$ gap between the $E_g$ and $A_{1g}$ peaks, aligning with findings from earlier studies.[13, 24]

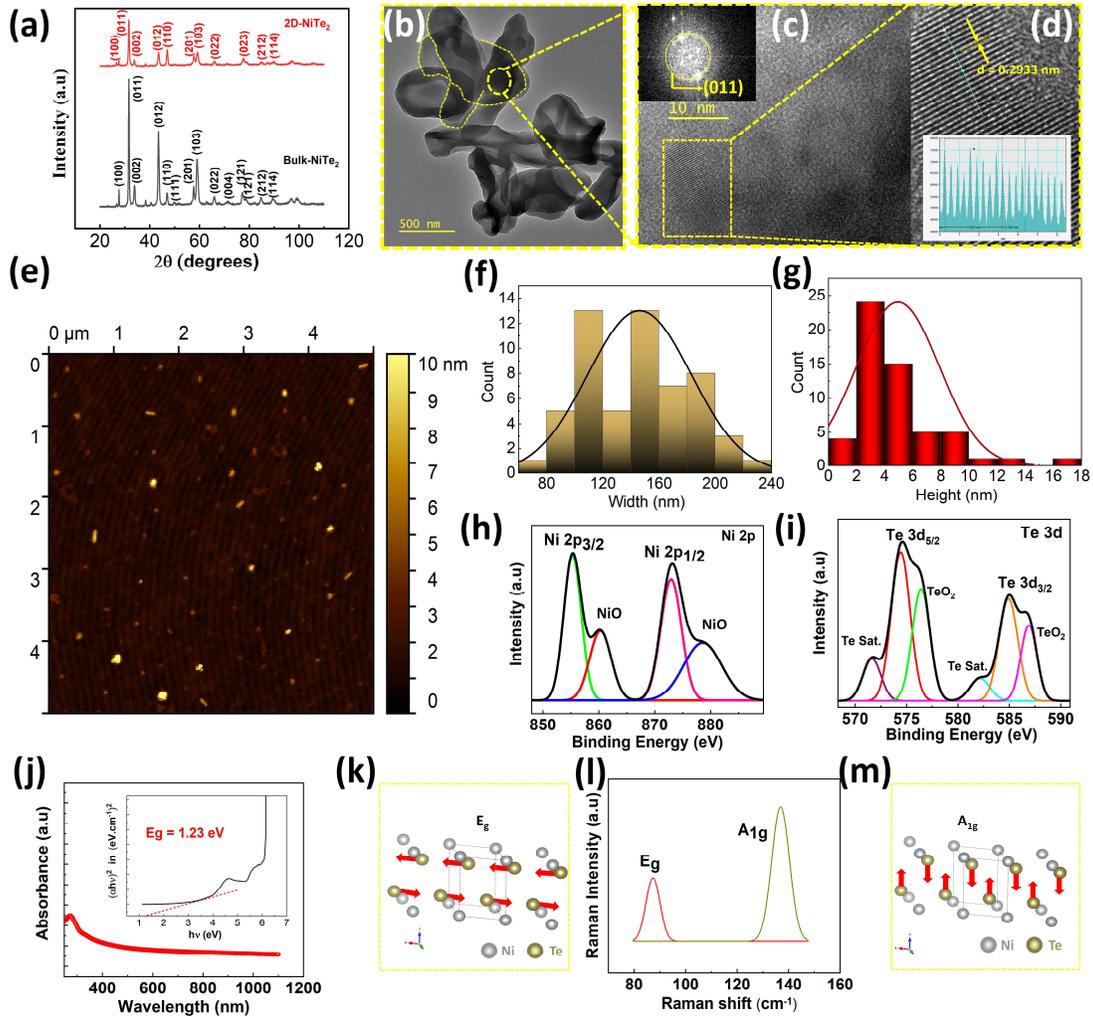

**Figure 1.** Characterization results of 2D-NiTe$_2$. a) Comparative XRD pattern analysis between Bulk and 2D. b) The HRTEM image of the exfoliated layers shows two separating layers in a dotted line. c) HRTEM image of the 2D-NiTe$_2$, inset showing FFT pattern of the (011) plane matched with the plane orientation. d) HRTEM image of the (011) plane along with the calculated d spacing value. The high-definition XPS spectrum of h) Ni 2p and i) Te 3d orbitals. e) AFM image of the flakes distributed on the surface. f-g) Showing the width (lateral dimension) and height (thickness) of the exfoliated flakes. j) The UV visible spectroscopy of the 2D-NiTe$_2$, inset showing Tauc plot for band gap estimation. l) Raman spectroscopy of 2D-NiTe$_2$ film. k-m) Schematic representation of one of E$_g$ and A$_{1g}$ mode, gray and golden sphere corresponds to Ni and Te atoms.

## 3. Results and Discussion

### 3.1. SSPM Spectroscopy Setup

**Figure 2a** shows a non-linear optical response characterization tool that was used for a spatial self-phase modulation (SSPM) experiment. Three lasers of different continuous wavelengths ($\lambda$ = 650, 532, and 405 nm) were used in this experiment. The laser beams were made to pass through the convex lens of a focal length of 20 cm and then onto the sample containing the cuvette. The SSPM effect becomes evident when the laser beam passes through the cuvette, generating a diffraction pattern on the far field screen, which is then recorded by a CCD camera. The fundamental principle underlying the nonlinear optical response of 2D materials based on SSPM can be elucidated through Kerr nonlinearity.

### 3.2. Nonlinear Kerr Effect: $n_2$ and $\chi^3_{total}$ Calculation

The Kerr effect plays a crucial role in examining the nonlinear optical characteristics of 2D material. The nonlinear refractive index can be stated by considering the nonlinear Kerr effect as,[27]

$$n = n_0 + n_2 I \qquad (2)$$

Here $n_0$ is the linear refractive index of the IPA solution, while $n_2$ signifies the nonlinear refractive index of the 2D material. The 2D-NiTe$_2$-IPA solution shows a strong Kerr nonlinearity response. A phase shift occurs as a laser beam passes through the solution with dispersed 2D material, leading to the observation of nonlinear modulating behavior in the traverse intensity profile of the transmitted laser beam.[18i] The expression for this phase shift in the beam profile can be stated as,[28]

$$\Delta\psi = \frac{2\pi n_0}{\lambda} \int_0^{L_{eff}} n_2 I(r,z) dz \qquad (3)$$

Where $\lambda$ is the wavelength of the incident laser beam and $L_{eff}$ represents the effective optical length of the dispersion medium containing 2D-NiTe$_2$ in the cuvette, $r \in [0, +\infty)$ is the radial coordinate and $I(r,z)$ is the radial intensity distribution. Changes in the intensity pattern of the incident laser beam can impact the modulation of the phase shift $\Delta\psi$. In the case of a beam with a Gaussian profile, the intensity at the center $I(0,z)$ can be expressed as $2I$, where $I$ denotes the average intensity. For the outgoing beam, there are at least two points (r$_1$ and r$_2$), where the Gaussian profile has two same slopes $(d\Delta\psi/dr)_{r=r_1}$ and $(d\Delta\psi/dr)_{r=r_2}$, and they exhibit the same phase difference, demonstrated in Figure 2b. Hence, the output light intensity profile exhibits a consistent phase difference when the slope points remain the same.

The formation of the ring pattern is governed by the following condition,[17]

$$\Delta\psi_{r_1} - \Delta\psi_{r_2} = 2N\pi \qquad (4)$$

Here N represents an integer, dark and bright field correspond to the odd and even values of N. The applicable length of the cuvette for the incoming laser beam can be articulated as,[29]

$$L_{eff} = \int_{L_1}^{L_2}(1 + \frac{z^2}{z_0^2})^{-1} \, dz = z_0 \arctan\left(\frac{z}{z_0}\right)\Big|_{L_1}^{L_2}, \; z_0 = \frac{\pi\omega_0^2}{\lambda} \quad (5)$$

In this context, $L_1$ is the distance from the back surface of the cuvette to the focal point and $L_2$ is the distance from the front cuvette surface to the focal point. The transmitted beam intensity profile is Gaussian in nature, the transmitted laser intensity beam profile can be stated as $I(0, z) = 2I$, here $I$ represent the mean intensity of the incoming laser beam, $z_0$ is the diffraction length and $\omega_0$ denotes the beam radius ($\frac{1}{e^2}$). From Equation 2 and 5, the $n_2$ can be written as,

$$n_2 = \frac{\lambda}{2n_0 L_{eff}} \frac{dN}{dI} \quad (6)$$

The $\frac{dN}{dI}$ parameter is important in calculating the NLO parameters in SSPM spectroscopy, the value of the parameter $\frac{dN}{dI}$ can be determined from the diffraction pattern. The third nonlinear susceptibility $\chi_{total}^{(3)}$ is defined as,[18e, 30]

$$\chi_{total}^{(3)} = \frac{cn_0^2}{12\pi^2} 10^{-7} n_2 \; (e.s.u) \quad (7)$$

In this context, $c$ represents the speed of light in free space, $n_0$ is the linear refractive index of the IPA solvent, the wavelength of the incident laser beam is denoted as $\lambda$, and $L_{eff}$ denotes the effective length that the laser beam propagates through the cuvette. The $\chi_{total}^{(3)}$ value is notably affected by the number of operative layers of the 2D-NiTe$_2$ nanostructure in the solvent. Therefore, it is essential to compute the value of the third-order nonlinear susceptibility for a monolayer $\chi_{monolayer}^{(3)}$. The correlation involving the strength of the electric field caused by the incoming laser beam ($E_{total}$) and the electric field strength passing through a monolayer ($E_{monolayer}$) can be related as $E_{total} = \sum_{j=1}^{N_{eff}} E_j \cong N_{eff} E_{monolayer}$.[17, 31] Here, $N_{eff}$ denotes the effective number of 2D-NiTe$_2$ layers within the solution along the path of the beam. The relationship between $\chi_{total}^{(3)}$ and $\chi_{monolayer}^{(3)}$ can be expressed as $\chi_{total}^{(3)} = N_{eff}^2 \chi_{monolayer}^{(3)}$.[32] The effective number of 2D-NiTe$_2$ present in the incoming beam pathway is calculated in the supporting information (Section S1).

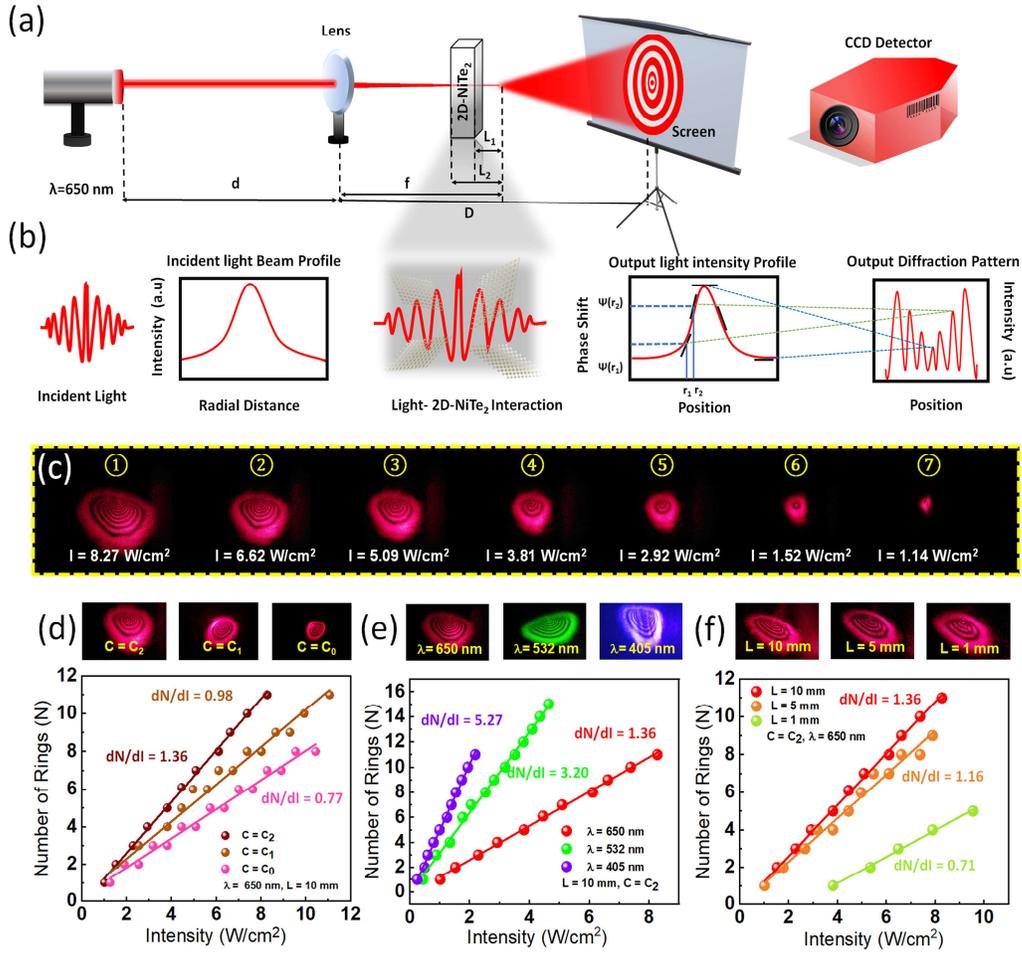

**Figure 2.** Experimental setup and SSPM spectroscopy-based results. a) Diagram illustrating the SSPM experiment. b) Schematic representation of 2D-NiTe$_2$-light matter interaction and the SSPM phenomenon by electronic coherent involvement with the laser beam. c) The progression of the diffraction pattern with the change in the intensity on the far field, for wavelength 650 nm, concentration C$_2$, and cuvette length 10 mm. d) The change in the number of diffraction rings vs varying laser (650 nm) intensity for different concentrations. e) The variation of the number of diffraction rings vs changing laser intensity for different wavelengths (650, 532, and 405 nm). f) The variation of the number of diffraction rings vs intensity for different cuvette lengths (10 mm, 5 mm, and 1 mm).

Figure 2c (①-⑦) displays the far-field SSPM pattern at a wavelength of 650 nm, captured by a CCD camera. A linear growth in the number of rings is noted as the intensity increases. The increase in the intensity causes the horizontal and vertical diameter of the diffraction pattern to increase. The same experiment has been demonstrated for two other different wavelengths (532 and 405 nm), shown in supporting information Figure S1 and S2. Figure 2e displays a diagram showing the correlation between intensity and the number of rings for the laser beams of different wavelengths (λ= 650, 532, and 405 nm). And $\frac{dN}{dI}$ is calculated from the linear fitted

data, estimated to be 1.36 cm$^2$W$^{-1}$, 3.20 cm$^2$W$^{-1}$, and 5.27 cm$^2$W$^{-1}$ for the wavelengths λ= 650, 532, and 405 nm respectively. In Supporting Information Section S6 the rationale behind using this particular wavelengths is discussed. According to the calculated data, it was understood that $\frac{dN}{dI}$ increases with the decreasing wavelength of the incoming laser beam. It was deduced that as the wavelength decreases, photon energy increases, resulting in a heightened SSPM effect. This conclusion aligns with the findings reported in earlier studies.[18h, 31] The relative position between the sample and the focal point of the lens in the SSPM experimental setup is shown in the Supporting Information Table S6. The SSPM pattern formation is also dependent on additional factors like, the effective presence of 2D material in the solvent and the effective path length of the incident laser beam within the cuvette. Figure 2d illustrates the correlation between the intensity of laser light with a wavelength of 650 nm and the number of rings with varying the concentration of 2D-NiTe$_2$ in the IPA solvent. This relationship is seen when examining different concentrations of suspended 2D materials in the solution. Curve fitting analysis revealed a linear correlation between the intensity of the laser and the number of diffraction rings. SSPM spectroscopy was performed at particular concentrations 0.0625 (C$_0$), 0.13 (C$_1$), and 0.25 (C$_2$) mg mL$^{-1}$ respectively, the associated slopes $\frac{dN}{dI}$ are found to be 0.77, 0.98, and 1.36 cm$^2$W$^{-1}$. Figure 2f illustrates the change in the number of rings relative to incoming laser beam intensity for different cuvette lengths. It was observed that the change in the number of diffraction rings with respect to the intensity ($\frac{dN}{dI}$) increases with the increasing the cuvette length, slopes are calculated to be 1.36, 1.16, and 0.71 cm$^2$W$^{-1}$ for cuvette length 10 mm, 5 mm, and 1 mm at a constant concentration of 0.25 (C$_2$) mg mL$^{-1}$. The sudden decrease in the number of rings at high intensity for 1 mm cuvette length can be attributed to the decreased light matter interaction. To verify above statement 2 mm cuvette length has been used to quantify the inetraction process, this is available in the Supporting Information Section S7. As the effective number of 2D-NiTe$_2$ decreases in the laser propagation path the value of ($\frac{dN}{dI}$) decreases. Through the experiment, it was found that the value of the $n_2$ for 2D-NiTe$_2$ are 3.22×10$^{-5}$, 6.15×10$^{-5}$, and 7.68×10$^{-5}$ cm$^2$W$^{-1}$ at the specified wavelengths of 650, 532, and 405 nm. The third order nonlinear susceptibility ($\chi^{(3)}_{total}$) is found to be 1.56×10$^{-3}$, 2.99×10$^{-3}$, and 3.76×10$^{-3}$ e.s.u. respectively at the wavelengths of 650, 532, and 405 nm. It was observed that the number of the rings changes with respect to intensity, and rises with an increase in both cuvette thickness and the concentration of suspended 2D-NiTe$_2$ within the cuvette. The calculated values of $n_2$ and $\chi^{(3)}_{total}$ from the above experiment is presented in **Table 1**. It was concluded from the experiment that the values of $n_2$ and $\chi^{(3)}_{total}$ increases by increasing the energy of the laser beam. The values of $n_2$ and $\chi^{(3)}_{total}$ increases with reducing the length of travel inside the cuvette and with the increase of the active material in the cuvette. The computed third order nonlinear susceptibility value for a monolayer ($\chi^{(3)}_{monolayer}$) exhibits dependence on the thickness of the cuvette and the concentration of the active nanomaterials. The obtained value of $n_2$ and $\chi^{(3)}_{total}$ for 2D-NiTe$_2$ through the SSPM method was found to be high compared to other transition metal dichalcogenides TMDCs (compared in between Table 1 and supporting information Table S1). A detailed calculation is done to determine the value of $n_2$, $\chi^{(3)}_{total}$, and $\chi^{(3)}_{monolayer}$ for the experiment done above. A comprehensive study is given in Supporting Information S5 to differentiate electronic coherence from the thermal lens effect, and evidence is included to demonstrate that the SSPM phenomena in 2D-NiTe$_2$ arises from the electronic coherence effect.

**Table 1.** Experimental values of $n_2$ (nonlinear refractive index), $\chi^{(3)}_{total}$ (third-order nonlinear susceptibility), and $\chi^{(3)}_{monolayer}$.

| Wavelength (nm) | Concentration (mg mL$^{-1}$) | L (mm) | dN/dI (cm$^2$ W$^{-1}$) | $n_2$ (cm$^2$ W$^{-1}$) | $\chi^{(3)}_{total}$ (e.s.u) | $\chi^{(3)}_{monolayer}$ (e.s.u) |
|---|---|---|---|---|---|---|
| 650 | 0.25 | 10 | 1.36 | $3.22 \times 10^{-5}$ | $1.56 \times 10^{-3}$ | $4.06 \times 10^{-9}$ |
| 532 | 0.25 | 10 | 3.20 | $6.15 \times 10^{-5}$ | $2.99 \times 10^{-3}$ | $7.79 \times 10^{-9}$ |
| 405 | 0.25 | 10 | 5.27 | $7.68 \times 10^{-5}$ | $3.76 \times 10^{-3}$ | $9.8 \times 10^{-9}$ |
| 650 | 0.13 | 10 | 0.98 | $2.34 \times 10^{-5}$ | $1.13 \times 10^{-3}$ | $1.09 \times 10^{-8}$ |
| 650 | 0.0625 | 10 | 0.77 | $1.82 \times 10^{-5}$ | $0.88 \times 10^{-3}$ | $3.66 \times 10^{-8}$ |
| 650 | 0.25 | 5 | 1.16 | $5.5 \times 10^{-5}$ | $2.65 \times 10^{-3}$ | $2.762 \times 10^{-8}$ |
| 650 | 0.25 | 1 | 0.71 | $1.68 \times 10^{-4}$ | $8.13 \times 10^{-3}$ | $2.11405 \times 10^{-6}$ |

For the SSPM pattern resulting from the thermal lens effect, the $\chi^{(3)}_{total}$ value falls within the $10^{-7}$ e.s.u. range and $n_2$ in the range of $10^{-9}$ cm$^2$W$^{-1}$.[33] In this experiment, the laser intensity is kept to a minimum to avoid this thermal lens effect.[33] Considering relevant literature[34], existing Dirac-like cone near the Fermi level for NiTe$_2$, may give rise to laser-induced hole coherence. It is to be noted that the valence band was found to lie above the Fermi level, exhibiting tilted band crossing, and the material is found to be type-II Dirac semimetal. The presence of the Dirac-like cone near the fermi level is likely to lead to more substantial contributions from these relativistic carriers in the materials transport characteristics.[7a, 35] This proximity causes the charge carriers to have topological properties, which ultimately makes nonradiative recombination to be effortless compared to the radiative recombination. Due to carrier-phonon scattering, the carriers will go to advancement in the momentum space and fall into the Dirac cone, and recombine with their counterpart. Cheng et al. noted that the rapid relaxation process cannot be explained by electron-phonon scattering. However, the authors rejected the idea of radiative recombination through a direct bandgap as the cause.[36] The authors concluded that the tilted positioning of Dirac cones near the valence band maxima enhances the phonon-assisted electron-hole recombination, which is a topological signature of the NiTe$_2$. This phonon-assisted electron-hole recombination process can be explained by the accumulation of the holes after the electron-phonon thermalization process. The carriers falling into these Dirac-like cones will contribute to the SSPM pattern formation. Due to the cumulative effect of these topological charge carriers in the Dirac cone, carriers will contribute to the large value of $\chi^{(3)}_{monolayer}$ as calculated above. Carriers that recombine through the Dirac-like cones also have the chance of contributing to the SSPM formation, as the carriers can get affected by light field easily.[34a] During this recombination process, the carrier adopts the properties, like high hole mobility. This causes the high value of the $\chi^{(3)}_{monolayer}$. Density functional theory calculation is included towards the end of the paper to support the following claim of the Dirac cones, also discuss the contribution of the Te-5p orbital in hole induced charge transport. Further mechanism enhancing NLO behavior due to Dirac Semi metal is added in the Supporting Information Section S9.

### 3.3. SSPM Diffraction Pattern Formation

Here, we examine the process of pattern generation or the underlying mechanism of the SSPM. In ref [18e] Wu et al. proposed a model to explain the process of pattern development in the SSPM process, which includes the time needed for the pattern to form at a constant intensity. Based on this wind chime model, the photoexcited carriers (either holes or electrons) in each 2D-NiTe$_2$ are influenced by the incoming laser beam, increasing electron/hole coherence.[31] Each 2D nanostructure is considered to be a distinct domain, and they are oriented in various orientations. The laser beam causes polarization in the 2D nanostructure suspended in the solvent. Subsequently, the polarized 2D-NiTe$_2$ layers or the domain axis undergo reorientation in accordance with the orientation of the electric field, generated by the laser beam via an energy relaxation process.[18e] Once all the domains are aligned nonlocal electronic coherence is reached, and the maximum number of rings will be formed. **Figure 3a** explains the wind chime model for 2D-NiTe$_2$ in the IPA dispersion. Figure 3b (①-⑦), 3c (①-⑦), and 3d (①-⑦) depict the progression of the SSPM pattern under the influence of continuous-wave laser beams of different wavelengths λ = 650, 532, and 405 nm. At first, a bright spot appears in the far field, and over time, the whole diffraction process occurs when 2D-NiTe$_2$ layers or domains align more precisely with the electric field. The duration it takes for the 2D-NiTe$_2$ layers or domains to align completely with the electric field of the incident laser beam is equivalent to the time taken for the full-diameter diffraction ring pattern to form. The ring formation process follows the exponential model[18g], as $N = N_{max}\left(1 - e^{t/\tau_{rise}}\right)$. Here, $N$ represents the number of rings in the diffracted pattern, $N_{max}$ is the maximum number of rings that can be formed under constant laser intensity, and the $t_{rise}$ is the rise time for the pattern formation. The duration needed for the emergence of full diameter diffraction beam can be accurately represented by the wind chime model as,[18e, 18h]

$$\mathcal{T} = \frac{\epsilon_r \pi \eta \xi R_C}{1.72(\epsilon_r - 1)Ih} \tag{8}$$

here $\varepsilon_r$ represents the relative dielectric constant of NiTe$_2$ understood to be 8.88 [7b], η represents the viscosity coefficient of the solvent (2.37×10$^{-3}$ Pa.s for at 20°C), $R_C$ is the radius of the 2D-NiTe$_2$ nanostructure with its exposed surface to the solvent (domain), h is the vertical height of the suspended 2D flakes, and $I$ is the incident laser beam intensity. Through atomic force microscopy, the determined value of domain radius (R) was found to be 0.145 µm, and the flake thickness (h) is 5 nm. Additionally, the value of $\mathcal{T}$ is estimated to be 0.172 s, 0.305 s, and 0.586 s for the wavelengths 650, 532, and 405 nm using Wind Chime Model. That is consistent with experimentally calculated values of the time taken to reach maximum number of rings. Experimentally calculated values following the exponential equation model using curve fitting to be 0.102, 0.194, and 0.131 for the wavelengths 650, 532, and 405 nm, respectively, as shown in Figure 3 e-g. These values are consistent with previously reported values, as seen in supporting information Table S2.

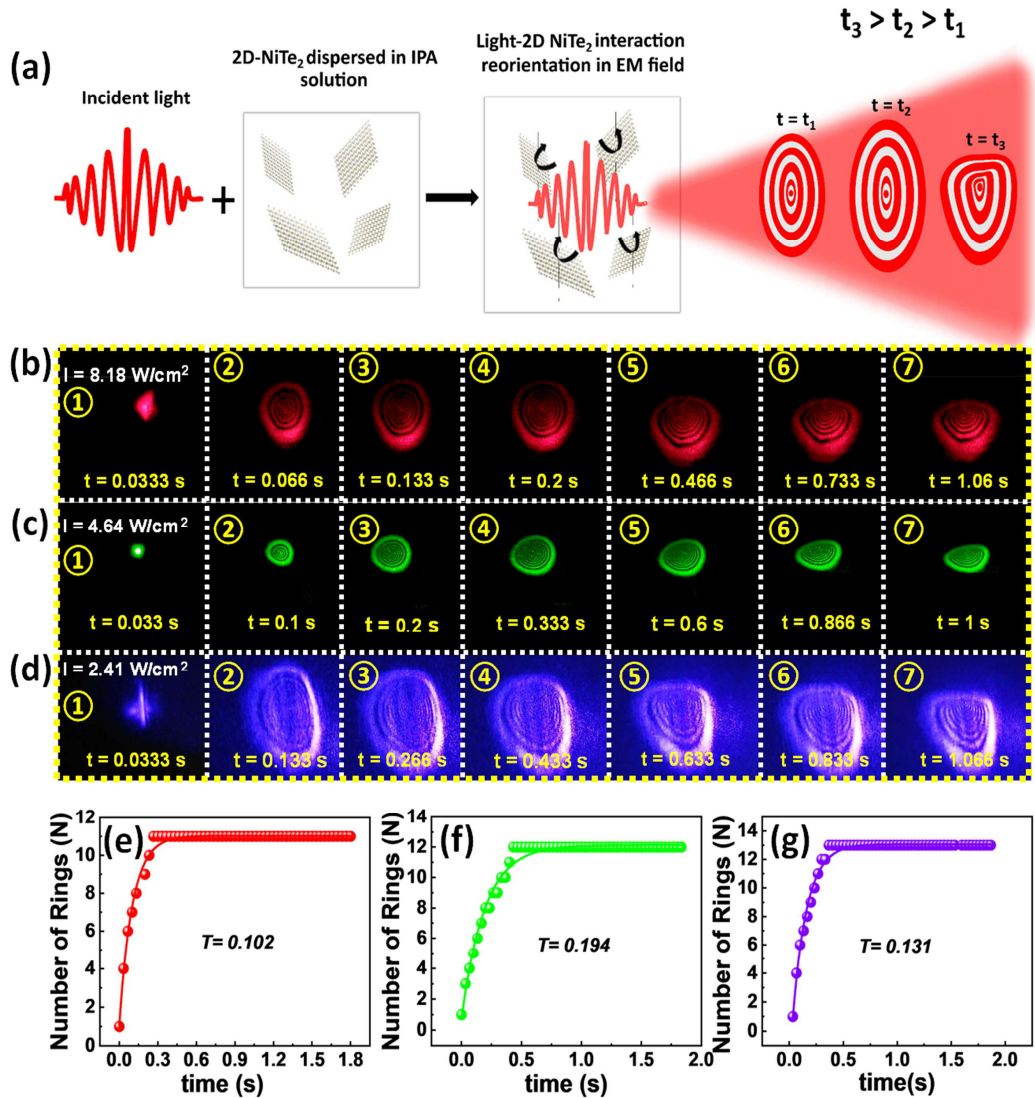

**Figure 3.** Demonstrating time evolution of the diffraction pattern of the 2D-NiTe$_2$ layers in the IPA solution. a) Schematic to present the working of the "wind chime" model to describe electronic coherence created due to incoming laser beam. b-d) The evolution of diffraction rings on the far field as a function of time for different wavelengths (650 nm, 532 nm, and 405 nm). e-g) The evolution of the SSPM pattern at high intensities with time for different wavelengths (650 nm, 532 nm, and 405 nm).

### 3.4. Dynamic Collapse of Diffraction Ring and Variation in Nonlinear Refractive Index

In the context of the SSPM effect, the diffraction pattern gradually appears when the laser beam passes through the nonlinear medium, namely the cuvette containing 2D-NiTe$_2$ in IPA solution. After observation, it was revealed that the diffracted pattern is progressively being deformed over time. When the diffraction pattern achieves the highest number of rings, the top section of the pattern starts converging towards the central section of the diffraction pattern. Distortion is more pronounced in the vertical direction compared to the horizontal direction. This distortion of the pattern can be described by non-axis thermal convection, first reported by Wang et al.[27], explains the phenomenon behind the collapse process. The maximum radius and the corresponding half-cone angle are considered to be $R_H$ and $\theta_H$.[18i] The relation can be expressed as,

$$\theta_H = \frac{R_H}{D} \qquad (9)$$

if $D \gg R_H$, D represents the distance between the cuvette and the screen. After some time, due to the distortion caused by the thermal convection effect, the maximum diffraction radius and the half-cone angle are changed to $R_h$ and $\theta_h$. If $D \gg R_h$, then the relation can be expressed as,

$$\theta_h = \frac{R_h}{D} \qquad (10)$$

Therefore, the distortion is quantified as distortion radius and the distortion angle is as follows, $R_D$ and $\theta_D$. The relation can be described as,

$$\theta_H - \theta_h = \frac{R_D}{D} \qquad (11)$$

In addition, the half-cone angle can be estimated as,

$$\theta_H = \frac{\lambda}{2\pi}\left(\frac{d\psi}{dr}\right)_{max}, r \in [0, +\infty] \qquad (12)$$

Here r represents the transverse position of the beam. For a Gaussian beam, the above Equation can be $\theta_D$ can also be expressed as,[37]

$$\theta_H = n_2 I C \qquad (13)$$

Where $C = \left[-\frac{8rL_{eff}}{\omega_0^2}\exp\left(-\frac{2r^2}{\omega_0^2}\right)\right]$ is constant when $r \in [0, +\infty)$. Consequently, we can articulate the distortion angle. **Figure 4a** shows the change in the diffraction pattern with time. This variation in distortion can be measured through an intensity-dependent examination of the angle of distortion formed between the sample and the screen.[18h] Therefore, we can get the collapse angle as,

$$\theta_D = \theta_H - \theta_h = (n_2 - n_2')IC = \Delta n_2 IC \qquad (14)$$

The distortion exhibited in the upper portion of the diffraction pattern is caused by the heat-driven fluid motion that lacks axial symmetry, leading to the collapse of the diffraction pattern.[27] The thermal convection arises from the propagating laser beam, solvent heats up due to its limiting value of absorption coefficient.[18h, 38] For this, a temperature gradient is generated perpendicular to the laser point's axis, augmenting the 'thermal convection process'. From Equation 13 and Equation 14 we can conclude that,

$$\frac{\Delta n_2}{n_2} = \frac{\theta_D}{\theta_H} = \frac{R_D}{R_H} \qquad (15)$$

Depending on the Equation, the distortion process can be concluded with the term $\Delta n_2/n_2$ that signifies a dynamic change in distortion and full radius of diffraction rings ($R_D/R_H$). The value of $\Delta n_2/n_2$ primarily depends on laser intensity.[37, 39] The term $\Delta n_2/n_2$ is affected by parameters such as laser wavelength, temperature, and time. The value of $\Delta n_2/n_2$ can be obtained through calculation. The 532 nm laser induces high distortion compared to the 650 nm laser at a similar intensity as, each photon of the 532 nm laser beam contains more energy than the 650 nm laser. The distortion caused by the blue light tends to be the highest. However, the increase in distortion saturates after a particular intensity of the incoming laser beam.

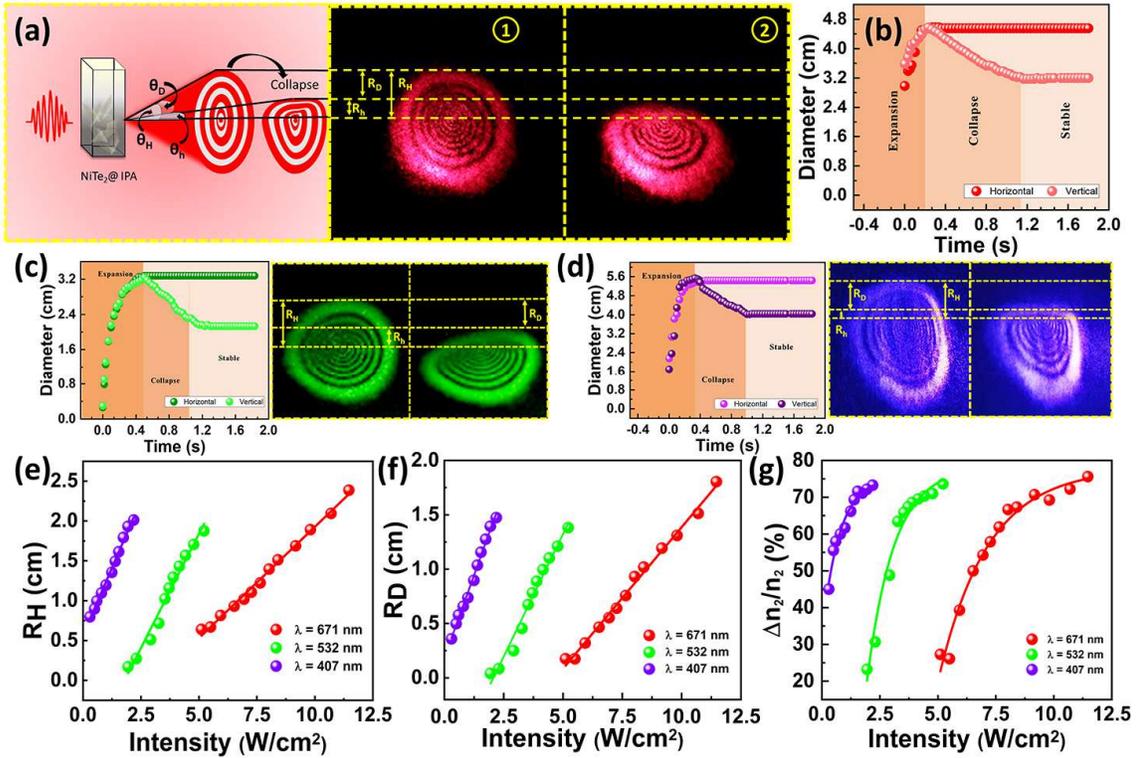

**Figure 4.** Visual representation of the dynamic collapse process. a) Diagrammatic depiction of the collapse phenomenon of the SSPM ring with a semi-cone and distortion angle. b-d) The evolution of the vertical and horizontal diameter with time for different wavelengths (650, 532, and 405 nm). e) The alteration in the maximum radius ($R_H$). f) Variations in the distorted radius ($R_D$) of the diffraction pattern with distinct laser wavelengths of λ = 650 nm, 532 nm, and 405 nm. g) The variation in the nonlinear refractive index concerning changing intensity with different wavelengths (650, 532, and 405 nm).

The experimentally calculated values of maximum radius before the collapse ($R_H$) and after the collapse ($R_h$). The collapse time of the diffraction pattern decreases as the photon's energy is increased, quantified as 0.933 s, 0.8 s, and 0.6667 s for the wavelength 650, 532, and 405 nm, respectively.

**Figure 4a** (①,②) shows the snapshot taken by the CCD camera. The diffraction pattern shows a maximum number of rings with the highest vertical diameter, and the equilibrium state after the collapse phenomenon respectively, with the laser beam of wavelength λ= 650 nm. Figure 4 b-d demonstrates the incremental progression of the diffracted SSPM pattern with three different lasers of wavelength λ = 650, 532, and 405 nm at relatively high intensity. It was observed that the vertical diameter increased for a particular period and then started to collapse. Figure 4 e-f illustrates the experimentally identified maximum vertical radius during pre-collapse and post-collapse, respectively. Due to the thermal convection process, a relative change in the SSPM diffraction ring was observed. This distortion is quantified with a variation of the nonlinear refractive index, following the Equation 15. The distortion $\Delta n_2/n_2$ rises with the increasing value of laser intensity as shown in Figure 4g. The distortion or relative nonlinear refractive index was found to be 75% (650 nm), 74.53 % (532 nm), and 73.26 % (405 nm) for intensities 11.46 Wcm$^{-2}$, 5.17 Wcm$^{-2}$, and 2.18 Wcm$^{-2}$. The expansion and collapse times are re-assessed using a 532 nm film polariser at an intensity of 5.80 Wcm$^{-2}$. This experiment is reported in Supporting Information Section S3.

### 3.5. 2D NiTe$_2$ – 2D hBN Based Nonlinear All-Optical Diode

A novel photonic nonlinear diode has been realized using 2D-NiTe$_2$/2D-hBN hybrid structure through SSPM spectroscopy. The material, 2D-hBN is prepared using LPE method and has a high optical bandgap (3.24 eV) compared to 2D-NiTe$_2$ (documentation of the 2D-hBN (Two-dimensional hexagonal boron nitride) characterization is available in the supporting information (Section S2)). Hence this heterostructure of 2D-NiTe$_2$/2D-hBN can be used in all-optical photonic diode application.[31, 40] The laser lights of wavelength 650, 532, and 405 nm have been used to implement the photonic diode application through SSPM. The rationale behind selection of these particular wavelengths 650, 532, and 405 nm is discussed in Supporting Information Section S8. In this experiment, the separate cuvettes containing 2D-hBN and 2D-NiTe$_2$ have a common thickness of 10 mm and both solution has a concentration of 0.25 mg.mL$^{-1}$. If the arrangement is 2D-NiTe$_2$/2D-hBN as described in the **Figure 5a,** the SSPM pattern is formed, it can be viewed in the Figure 5c (①-⑤), 5d (①-⑤), and 5e (①-⑤) for λ=650, 532, and 405 nm. In the reverse arrangement, as seen in Figure 5b the 2D-hBN/2D-NiTe$_2$, the beam profile is always Gaussian, as the reverse saturation ability of the hBN results in the reduction of the transmitted laser intensity. The diffraction pattern is not generated by this low-intensity laser beam because it falls below a threshold level. This reverse saturable absorption type behavior is observed in the 2D-hBN material, which is exploited for the realization of this all-optical diode arrangement.[41] This gaussian beam profile can be observed in the Figure 5c (⑥-⑩), 5d (⑥-⑩), and 5e (⑥-⑩) for λ=650, 532, and 405 nm. The calculated values of $\frac{dN}{dI}$ for the wavelengths (650, 532, and 405 nm) are found to be 1.22, 2.96, and 5.06 cm$^2$W$^{-1}$, respectively for forward bias condition. These reported values are quite equivalent to the ones obtained from a single cuvette 2D-NiTe$_2$-IPA solution. The linear-fitted graphs for the forward bias condition and reverse bias condition are shown in Figure 5 f-h. Here, asymmetric light propagation is realized by using three different types of lasers, with different wavelengths of λ= 650, 532, and 405 nm. All these wavelengths have photon energy exceeding the bandgap of 2D-NiTe$_2$, which is 1.23 eV. Therefore, the laser photons could stimulate the band-to-band transition. For forward arrangement as shown in Figure 5i, the photons of the incoming laser beam excite the valence band electrons to make a transition to the conduction band. Then the energized electrons undergo a transition back to their ground state, releasing a photon in the process, this emitted photon then interacts with the incoming laser beam, resulting in the generation of a diffraction pattern.[18h, 40] The electrons in the conduction band will oscillate along the antiparallel axis to the electric field of the laser beam, which will cause a polarization of charges in the suspended material.[18e] These polarized flakes will reorient themselves along the axis of the electric field of the laser beam to achieve minimum interaction energy arrangement. This enhances the nonlinear optical response of 2D-NiTe$_2$, and the optical Kerr effect is observed.[18h, 37] For 2D hexagonal boron nitride (2D-hBN), the band-to-band transition does not occur by laser beam excitation of the wavelengths λ= 650, 532, and 405 nm. The electrons lose their energy while making an intraband transition. Due to the reverse saturable absorption property of 2D-hBN, the incoming beam intensity is reduced below a threshold such that no SSPM diffraction pattern is generated from the 2D-NiTe$_2$-IPA solution, as shown in Figure 5j. This all-optical diode, as claimed, has a broad operating wavelength range. Similar comparison method can be used to calculate the value of $n_2$ for NiTe$_2$ based photonic diode, discussed is Supporting Information Section S8. Compared to other diodes where magnetism, polarization is used to gain non-reciprocal behavior. To confirm this a optical setup with film polarizer is arranged to see any change in the nonlinear optical behavior. The above mentioned experiment is performed for wavelength 532 nm only, gained results are discussed in Supporting Section S4.

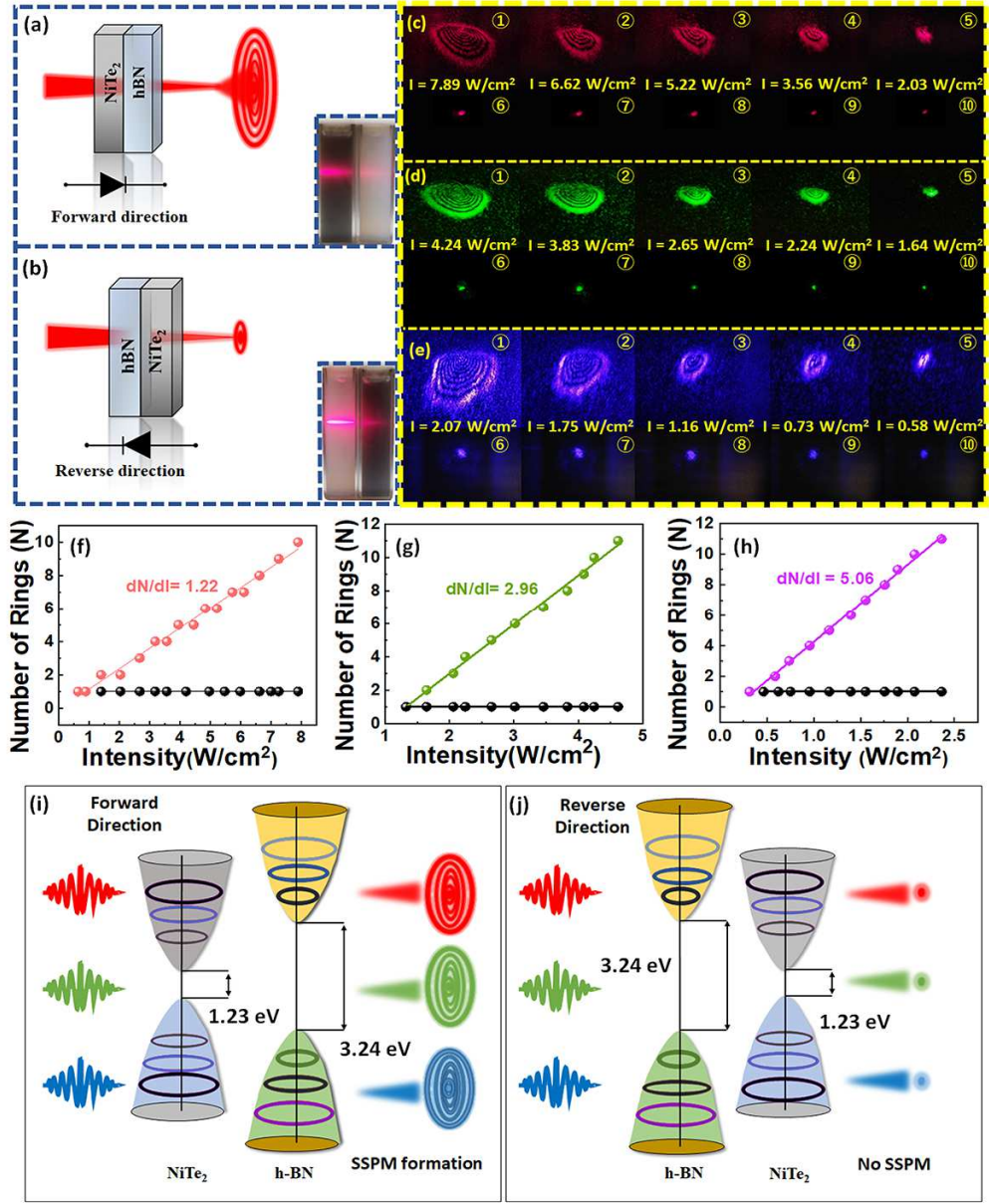

**Figure 5.** Realization of all-optical photonic diode. a-b) Figure showing forward and reverse biased analogy in case of the all-optical diode. c-d-e) (①-⑤) Shows the diffraction rings vs intensity for forward condition for different wavelengths (650 nm, 532 nm, and 405 nm). c-d-e) (⑥-⑩) Shows the diffraction rings vs intensity for reversed condition for different wavelengths (650 nm, 532 nm, and 405 nm). f-h) The experimental result obtained in the case of asymmetric light transmission light propagation for λ= 650, 532, and 405 nm in forward bias mode reverse bias mode. i-j) The diagram illustrates the band diagram for the functioning of an all-optical photonic diode in both forward and reverse directions.

## 3.6. Electronic Relation Between $\chi^{(3)}_{monolayer}$, Mobility ($\mu$) and Effective Mass ($m^*$)

Furthermore, to investigate the origin of the SSPM pattern formation, a relationship is established between $\chi^{(3)}_{monolayer}$ and mobility, $\chi^{(3)}_{monolayer}$ and effective mass. Hu et. al. proposed a method to correlate the electronic coherence phenomenon with the carrier mobility and the effective mass.[18g] Carrier mobility defines the ability of a carrier to move in response to an external electric field. This motion of the carrier in the ab plane of the unit cell is governed by the effective mass and the scattering of the carrier. High value of $\chi^{(3)}_{monolayer}$ results due to less scattering electrons in the 2D material. Hence, by this logic, the optical coefficient $\chi^{(3)}$ can be correlated with the electrical parameters like effective mass ($m^*$) and carrier mobility ($\mu$). In this experiment, we have correlated the value of $\chi^{(3)}_{monolayer}$ obtained by the SSPM experiment with the value of mobility and effective mass of NiTe2 found in the relevant literature. Here, these results show the existence of nonlocal electron coherence of 2D material during the SSPM experiment. The $\chi^{(3)}$ of other materials are found to be in the relevant literature, and the value of carrier mobility (μ) and effective mass ($m^*$) is obtained from Table S3. The calculated value of $\chi^{(3)}$ and its correlation to electronic activity is as per as the TaAs and Black Phosphorous, but less than Graphene, this is shown in **Figure 6a** and **6b,** where the curve shows that the relation between, $\chi^{(3)}$ vs $\mu$ and) $\chi^{(3)}$ vs $m^*$,[18g] i.e.

$$\chi^{(3)} = 8.00/\sqrt{m^*} \qquad (16)$$

And
$$\chi^{(3)} = 0.146 \times \sqrt{\mu} \qquad (17)$$

The large value of $\chi^{(3)}_{monolayer}$ was explained by the existence of the Dirac cones.

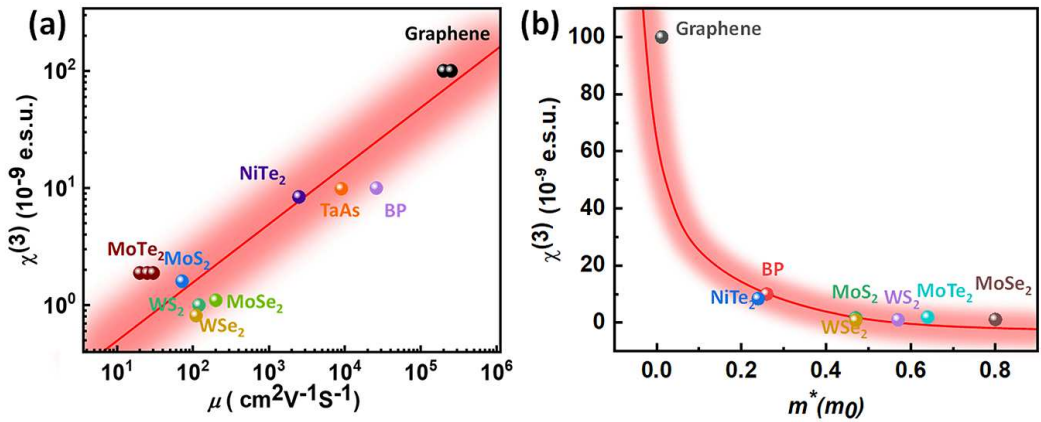

**Figure 6.** Comparative study between optical and electronic coefficients. a) $\chi^{(3)}_{monolayer}$ versus carrier mobility ($\mu$). b) $\chi^{(3)}_{monolayer}$ versus effective mass ($m^*$).

## 4. Density functional Theory Based Calculations

### 4.1. Computational Details

To gain an atomistic understanding of the electronic properties of bulk and 2D Nickel Telluride ($NiTe_2$), first-principles calculations based on Density Functional Theory (DFT) were carried out using the Quantum ESPRESSO open-suite.[42] The core electrons were treated within the frozen core approximation using the pseudopotentials method, obtained from the Standard Solid-State Pseudopotentials.[43] Khon-Sham orbitals were expanded using a plane wave basis set (as implemented in QE pwscf package) employing a cutoff of 45 Ry and 360 Ry for the kinetic energy and electronic density, respectively. For the bulk calculations, the Brillouin zone was sampled by a uniform grid following the scheme proposed by Monkhorst-Pack[44] with (24x24x24) 4x4x4 k-points for (non)self-consistent calculations. For the 2D structure, the k-point mesh employed was (24x12x1) 4x2x1, respectively. The exchange and correlation functions were approximated by the Perdew, Burke, and Ernzerhoff parametrization of the Generalized Gradient (GGA-PBE).[45] The convergence threshold for self-consistent calculations was set to $10^{-7}$ Ry. Energy and forces were minimized until the difference between two consecutive ionic steps was below $10^{-4}$ Ry and $10^{-3}$ Ry/Bohr, respectively. Since Ni poses d shell electrons, we have employed the Hubbard correction in the calculation of the electronic properties of bulk and 2D $NiTe_2$.

### 4.2. Structural Optimization

Bulk 1T-$NiTe_2$ poses a hexagonal lattice with $P\bar{3}m1$ space group symmetry. The structure is composed of an atomic layer of Ni octahedrally coordinated with Te atoms, forming a monolayer, as shown in **Figure 7a**. Long-range interactions ensure the interaction between the monolayer stacking and then into the characteristic TMD bulk structure. First, we optimize the Hubbard parameter (U), employing different values and comparing the resulting lattice parameters. For the initial guess, we use Density Functional Perturbation Theory (DFPT) with linear response calculations as implemented in the hp.x computational code of QE. The obtained U values and corresponding lattice parameters are shown in Table 3. The results show that the lattice parameters are closer to the experimental values for $U_{Ni}$ = 10.344 eV. For U = 9.207, 9.568, and 9.509, there is a small deviation for the lattice parameter while c is smaller. Specifically, in the fourth iteration, the a and c lattice parameters are overestimated by 5.5 % and 0.56%, respectively. Following these results, we use $U_{Ni}$ = 10.34 eV throughout this work. Since $NiTe_2$ has vdW interactions between the monolayers, we have also investigated the optimization of the lattice parameters using long-range dispersions (vdW) corrections for the exchange and correlation functional. As seen by the results presented in Table S4, the vdW functionals give a closer c parameter but tend to largely overestimate a lattice parameter. Therefore, we have employed GGA-PBE exchange and correlation function since it presents a smaller deviation from the experimental values.

**Table 3:** Resulting Hubbard parameters (U) and corresponding optimized lattice parameters obtained from the linear response calculations and experimental lattice parameters of bulk NiTe$_2$ for comparison.

| Iteration | U (eV) | a (A) | c (A) |
|---|---|---|---|
| 1 | 10.344 | 3.895 | 5.254 |
| 2 | 9.207 | 3.898 | 5.237 |
| 3 | 9.568 | 3.899 | 5.236 |
| 4 | 9.509 | 4.072 | 5.305 |
| Exp.[45] | - | 3.859 | 5.275 |

### 4.3. Structural Model for the 2D NiTe$_2$

The XRD experimental characterization reveals the presence of (011) surface of NiTe$_2$ after exfoliation. Following these results, we have constructed a 2D model by the cleavage of NiTe$_2$ in the (011) direction (as shown in Figure S4a. The resulting 2D monolayer (shown in Figure S4b) was used throughout the simulations. The (011) monolayer has triclinic symmetry with P1 space group. The optimized lattice parameters are a = 3.867Å and b = 6.853 Å. To avoid interactions between periodic images we have applied a 15 Å vacuum in the c direction. Since the (011) surface has uncoordinated Ni atoms, we have considered (non)magnetic states of the structure in an initial configuration where both Ni atoms have parallel magnetic moments (FEM) and the case where the magnetic moments are opposite (AFM) as represented in Figure S5. These magnetic moments were set as the initial magnetization of the Ni atoms and the final moments were calculated by QE on the optimization run. The results for the magnetic moments on both Ni atoms at the beginning and after geometry optimization are presented in Table 4. One can observe that in both cases, the optimized structures present zero magnetic moment and therefore the non-magnetic configuration is taken throughout the simulations.

**Table 4:** Results for the initial and final magnetic moments on the Ni atoms for the FEM and AFM configurations.

| Atom | FEM | | AFM | |
|---|---|---|---|---|
| | $\mu_i$ ($\mu_B$) | $\mu_f$ ($\mu_B$) | $\mu_i$ ($\mu_B$) | $\mu_f$ ($\mu_B$) |
| Ni$_1$ | 1.1760 | 0.0000 | 1.2288 | 0.0000 |
| *Ni₂* | 1.1759 | 0.0000 | -1.2286 | 0.0000 |

## 4.4. Electronic Properties of Bulk and 2D-NiTe₂

We now turn our attention to the electronic structure of bulk and 2D NiTe$_2$. Previous studies have reported that bulk 1T-NiTe$_2$ presents a semimetal nature with paramagnetic properties.[46] The electronic band structure for bulk NiTe$_2$ along the conventional path in the hexagonal lattice Brillouin zone is presented in Figure S6a. Indeed, the resulting magnetic moment calculated for Ni atoms with PBE+U is found to be 0.29 $\mu_B$, in good agreement with the results obtained by Aras and coworkers using PBE+U with spin-orbit coupling.[46c]

Mukherjee et al.[47] showed topological surface state and bulk type-II Dirac points of NiTe$_2$. Hlevyack et al.[48] proposed that as the number of layers increases, the gap in the topologic surface state closes in. According to the authors[48] as the layer number increases up to 4TL hole-like bands appear around Γ point, which is characteristic of the Dirac points as described by Ferreira et al..[49]

Type II Dirac -like cones are observed when a path is considered on the reciprocal space located at the intermediate plane in between the Gamma (Γ) and H high-symmetry points. The considered path is shown in **Figure 7b**. The position of the cones is identified in the electronic band structure diagram shown in Figure 7c at D symmetry points and in between the Γ-A path, as well as in the opposite direction in the momentum space. The valence band is observed to cross the E$_f$, the unique characteristic of the semimetal is characterized by a slight overlap between the conduction band and the valence band. These results in the presence of free electrons in the Ni-3d orbital. The topological behavior can be understood by discussing the symmetry distinguished Te$^1$ and Te$^2$ atoms and a single atom of Ni in the unit cell structure. The electronic configuration of Ni is $4s^23d^8$, and that of Te is $4d^{10}5p^4$. The Te 5p orbital manifold in NiTe$_2$, due to interlayer hopping and crystal field splitting, gives rise to most of the bulk Dirac nodes. NiTe$_2$ is a layered material, as the sheets are coupled along the direction of the c-axis, the fermi surface possesses a strong 3-dimensional character. The partial density of states (pDOS) is shown in Figure S6a. Analyzing the pDOS, a large asymmetry is observed between spin-up and down electronic states, evidencing the magnetic nature of the structure. It is observed the contribution of the Te-5p orbital is more compared to the Ni-d orbital. For instance, we have calculated the individual contribution of the electronic states in the energy range spanning [-40.0, 3.5] eV in the following manner: the pDOS for each individual orbital of each element was summed and divided by the sum of the total pDOS. According to this, the following results are obtained: Ni s, p, and d states represent 3.20%, 7.09%, and 35.08%, respectively, while Te s and p states contribute with 13.11% and 41.52%, respectively. This dominant contribution of Te 5p orbital causes the hole-enhanced conduction in the synthesized 2D-NiTe$_2$.[5b, 49]

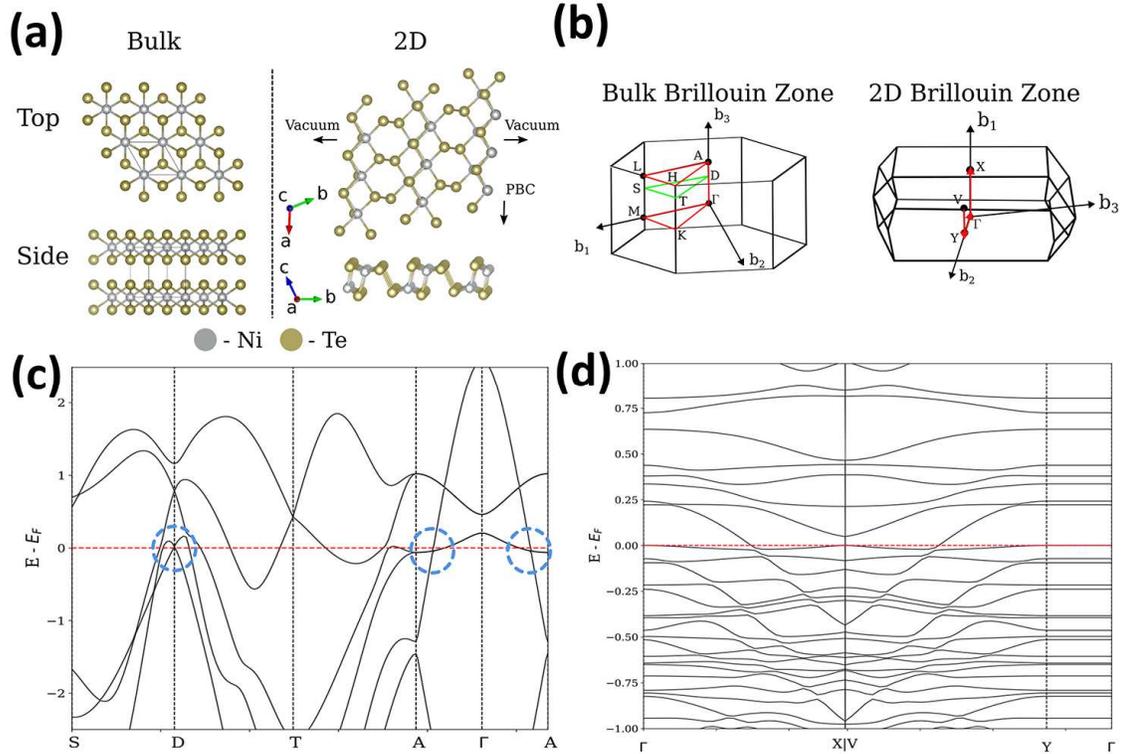

**Figure 7.** a) Top and side views of the crystal structure of Bulk 1T NiTe$_2$ and 2D-NiTe$_2$ (011) surface. b) The Brillouin zone (BZ) of Bulk NiTe$_2$ highlights the conventional hexagonal lattice path (red) and the path selected for the electronic structure (green) for the bulk, and the path selected along the BZ of the 2D structure. c) Electronic structure of bulk 1T-NiTe$_2$ along the non-conventional path (green) selected along the BZ. The blue dashed circles highlight the Dirac cones. d) Electronic structure for the 2D structure along the selected path.

We now analyze the electronic structure of the 2D-NiTe$_2$. The pDOS and electronic band structure for the (011) surface is shown in Figure S6b. The monolayer does not display semimetal behavior. For bulk, there are points that resemble Dirac-like cones in the vicinity of Z and Y points, although they are located at roughly –1.0 eV below the Fermi level. The pDOS shows that the electronic states have large contribution of Te-5p orbitals, especially in the Fermi level vicinity. To explore further reduction in the aspect ratio of the structure, we have built a model based on the 2D (011) NiTe$_2$ that is periodic only in the x direction (Figure 7a). The electronic band structure along the path in the reciprocal space (shown in Figure 7b right) is presented in Figure 7d. Reducing the dimensionality significantly alters the electronic structure.

The slight overlap between our data's valence and conduction bands suggests the semimetallic behavior is preserved, consistent with bulk-like properties. However, no Dirac-like features are observed in the 2D structure. This observation aligns with theoretical insights indicating that the thickness-dependent tuning of topological surface states (TSS) in NiTe$_2$ plays a crucial role[48]. Specifically, in ultrathin films, quantum tunneling between TSS on the top and bottom surfaces leads to hybridization, suppressing the net spin polarization and opening a hybridization gap. This effect becomes increasingly pronounced as the thickness decreases, ultimately resulting in the disappearance of the topological surface Dirac cone in the monolayer limit.

In our samples, AFM measurements show a thickness range of 5-10 nm, where bulk-like behavior dominates. The absence of Dirac-like features could therefore result from a combination of quantum tunneling effects and the presence of bulk-like properties in thicker films. Additionally, differences in sample preparation, such as substrate interactions or layer uniformity, could influence the electronic structure. These insights emphasize the critical role of dimensionality and interfacial effects in determining the topological behavior of NiTe$_2$.

## 5. Conclusion

In this study, we have synthesized high-quality 2D-NiTe$_2$ by a cost-effective liquid phase exfoliation method. Furthermore, we have elucidated the mechanism and physical origin behind the emergence of SSPM using 2D-NiTe$_2$ and explained the aforementioned phenomenon through the wind chime model considering electronic coherence. Their nonlinear refractive index ($n_2$) and third-order nonlinear susceptibility ($\chi^{(3)}_{total}$) is calculated, it was observed that the evaluated values are high compared to the reported literature. The nonlinear refractive index ($n_2$) is calculated to be 3.17×10$^{-5}$, 6.63×10$^{-5}$, and 7.68×10$^{-5}$ cm$^2$W$^{-1}$ for the wavelengths of 650, 532, and 405 nm. The third-order nonlinear susceptibility ($\chi^{(3)}_{total}$) is calculated to be 1.53×10$^{-3}$, 3.22×10$^{-3}$, and 3.76×10$^{-3}$ e.s.u. The high values of $n_2$ and $\chi^{(3)}_{total}$ are believed to result from laser-induced hole coherence. To further distinguish this effect from the thermal lens effect dependence between $\chi^{(3)}_{monolayer}$ vs $m^*$ and $\chi^{(3)}_{monolayer}$ and µ is derived, which resembles the relevant curves reported by other authors. A density functional theory (DFT) based method is utilized to analyze the band structure and density of states (DOS), providing an insight into the high values of ($\chi^{(3)}_{monolayer}$) and the contributing factor to this laser-induced hole coherence. As an application, SSPM-based photonic diodes are realized using 2D-NiTe$_2$/2D-hBN based heterostructure. The dN/dI derived from the 2D-NiTe$_2$/2D-hBN based heterostructure tends to align more closely with the dN/dI of 2D-NiTe$_2$; the proposed optical diode has optimal functionality. Our study can be viewed as a beneficial illustration of the realization of nonlinear, all-optical switching utilizing 2D-NiTe$_2$.

**Supporting Information**

Supporting Information is available from the Wiley Online Library or from the author.

**Acknowledgments**

S. Goswami and C. C. Oliveira contributed equally to this work. C.S.T. acknowledges DAE Young Scientist Research Award (DAEYSRA), and AOARD (Asian Office of Aerospace Research and Development) grant no. FA2386-21-1-4014, and Naval Research Board for funding support. C.S.T. acknowledges the funding support of AMT and Energy & Water Technologies of TMD Division of DST.A.S.A., C.C.O. and B.I thank CNPq (Grant 308428/2022-6), CNPq - INCT (National Institute of Science and Technology on Materials Informatics, grant no. 371610/2023-0), CAPES (finance code 001), the São Paulo Research Foundation (FAPESP process number #2024/11376-6), the UFABC Multiuser Computational Center (CCM-UFABC), the National Laboratory for Scientific Computing (LNCC/MCTI, Brazil) and Centro Nacional de Processamento de Alto Desempenho em São Paulo (CENAPAD-SP, Brazil) for providing HPC resources used in this work.

**Conflict of Interest Statement**

The authors declare no conflict of interest

**Data Availability Statement**

The data that support the findings of this study are available in the Supporting Information of this article.

**A diode that rectifies light:** The 2D-NiTe$_2$ was analyzed using SSPM spectroscopy to investigate its nonlinear optical characteristics. Key parameters such as the nonlinear refractive index and third order nonlinear susceptibility were computed. The discussion focuses on the evolutionary and distorted nature of the diffraction pattern of the SSPM. A nonlinear photonic diode is created utilizing the SSPM spectroscopy approach.

Saswata Goswami, Caique Campos de Oliveira, Bruno Ipaves, Preeti Lata Mahapatra, Varinder Pal, Suman Sarkar, Pedro A. S. Autreto*, Samit K. Ray* and Chandra Sekhar Tiwary*

**Exceptionally High Nonlinear Optical Response in Two-dimensional Type II Dirac Semimetal Nickel di-Telluride (NiTe₂)**

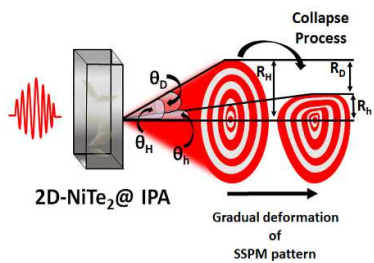

**Exceptionally High Nonlinear Optical Response in Two-dimensional Type II Dirac Semimetal Nickel di-Telluride (NiTe$_2$)**

Saswata Goswami, Caique Campos de Oliveira, Bruno Ipaves, Preeti Lata Mahapatra, Varinder Pal, Suman Sarkar, Pedro A. S. Autreto*, Samit K. Ray* and Chandra Sekhar Tiwary*

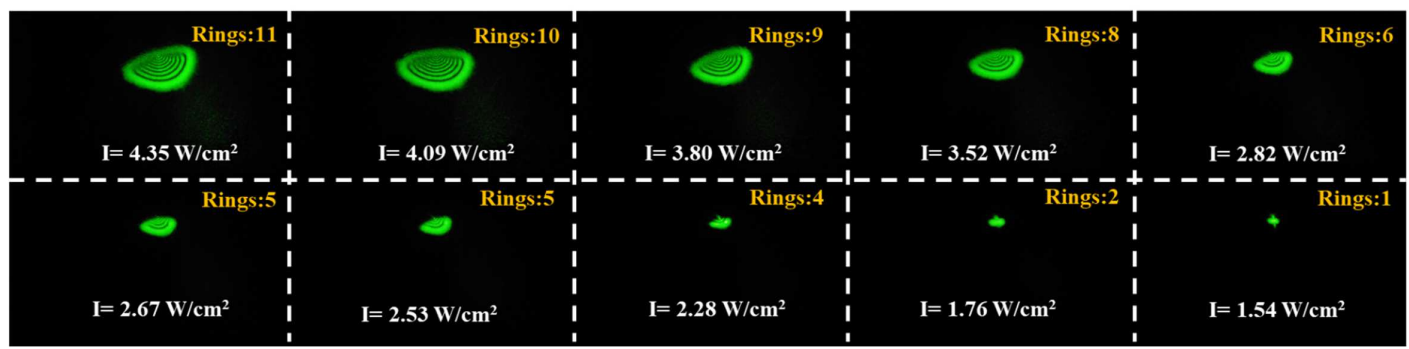

**Figure S1.** The progression of the diffraction pattern on the far field for the wavelength of 532 nm, concentration C$_2$ and cuvette length 10 mm

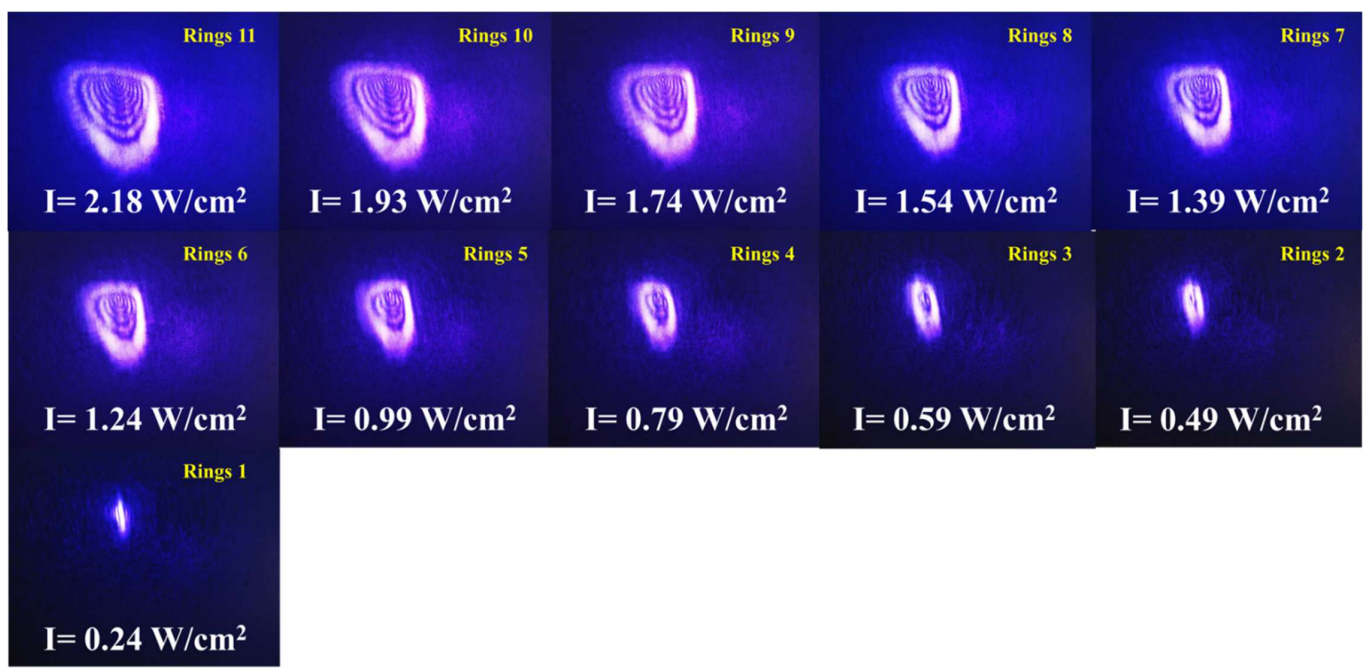

**Figure S2**. The progression of the diffraction pattern on the far field for the wavelength of 405 nm, concentration C$_2$ and cuvette length 10 mm.

**Section S1: Calculation of the effective number of the nanoflakes layers**

The concentration of 2D-NiTe$_2$ in the solution was varied, although 0.25 mg/ml solution was selected for the experimental purpose. The molecular weight of the NiTe$_2$ was found to be 313.89 g/mol. Hence the concentration can be expressed as 0.00079 mol/L. The volume of the cuvette is considered to be *4.5 × 10$^{-3}$ L*. Total number of molecules of NiTe$_2$ in the solution is M = $\rho \times V \times N_A$ ($N_A$ is the Avogadros constant). The space group of the NiTe$_2$ is $P\bar{3}m1$, which is a hexagonal crystal system, and lattice constants are found to be a = 3.869 Å and c = 5.30 Å, respectively. Thus, a single effective layer contains number of molecules *m = 1 × 4.5 cm$^2$ / (Sin 60°) × (3.869)$^2$ = 3.48 × 10$^{15}$* molecules[1]. The number of layers of the nanoflakes can be calculated as the following, n = $M/m$ = 619.

**Table S1.** The value of $n_2$ and $\chi^{(3)}_{total}$ calculated using SSPM Spectroscopy in recent literature.

| 2D Material | Type of laser | n$_2$ | χ$^3$ | χ$^3$ (monolayer) | References |
|---|---|---|---|---|---|
| Graphene | CW 532 nm | 2.5 X 10$^{-9}$ m$^2$/W | 1 X 10$^{-3}$ | 1 X 10$^{-7}$ | [2] |
| MoS$_2$ | CW 478 nm | 10$^{-7}$ | 1.44 X 10$^{-4}$ | 1.6 X 10$^{-9}$ | [1] |
| Ti$_3$C$_2$Ti$_x$ | CW 457 nm/ 532nm /671 nm | 11 X 10$^{-4}$/4.75 X 10$^{-4}$/4.72 X 10$^{-4}$ | | 4.34 X 10$^{-7}$/1.68 X 10$^{-7}$/0.15 X 10$^{-7}$ | [3] |
| 2D Te NS | CW 457 nm/ 532nm /671 nm | 6.14 X 10$^{-5}$/ 6.202 X 10$^{-5}$/ 7.37 X 10$^{-5}$ | ------ | ------- | [4] |
| Sb FS/QD | CW 532 nm / 633 nm | FS - 2.88 X 10$^{-5}$ / 0.979 X 10$^{-5}$ QD- 1.91 X 10$^{-5}$/ 0.719 X 10$^{-5}$ | FS - 3.98 X 10$^{-9}$ / 1.74 X 10$^{-9}$ QD- 2.87 X 10$^{-5}$/ 1.29 X 10$^{-5}$ | ----- | [5] |
| Bi$_2$Te$_3$ | CW 1070 nm | 2.91 X 10$^{-9}$ m$^2$/W | 10$^{-3}$ | 10$^{-8}$ | [6] |
| Graphene Oxide | CW 532 nm CW 671 nm | 3.57 X 10$^{-6}$ cm$^2$/W 1.1 X 10$^{-6}$ cm$^2$/W | 1.7 X 10$^{-6}$ 5.32 X 10$^{-6}$ | ------ | [7] |
| MoTe$_2$ | CW 473 nm /532 nm/ 750 nm/ 801 nm | ------ | ------- | 1.88 X 10$^{-9}$ esu 1.3 X 10$^{-9}$ esu 1.14 X 10$^{-9}$ esu 0.98 X 10$^{-9}$ esu | [8] |
| NbSe$_2$ | 532 nm / 671 nm / | 1.352 X 10$^{-5}$ m$^2$ W$^{-1}$ /2.0 X 10$^{-5}$ m$^2$ W$^{-1}$/1.07 X 10$^{-5}$ m$^2$ W$^{-1}$ | 1.352 X 10$^{-5}$ /9.354 X 10$^{-6}$ /5.03 X 10$^{-6}$ | 3.34 X 10$^{-9}$/2.59 X 10$^{-9}$ /3.39 X 10$^{-9}$ | [9] |
| Bi$_2$Se$_3$ | 350 nm/ 600nm/ 700nm/ 1160 nm | 1.16 X 10$^{-8}$ / 3.53 X 10$^{-9}$ / 2.5 X 10$^{-9}$/ 1.65 X 10$^{-9}$ (m$^2$/W) | 5.76 X 10$^{-3}$ /1.82 X 10$^{-3}$ / 1.29X 10$^{-3}$/8.53 X 10$^{-4}$ (esu) | 10$^{-8}$/10$^{-8}$/10$^{-9}$/10$^{-9}$ | [10] |
| Black Phosphorus | Pulsed Laser 350-1160 nm | 10$^{-5}$ cm$^2$W$^{-1}$ | 10$^{-8}$ esu | ----- | [11] |
| WSe$_2$ | CW 532nm /671 nm/457 nm/ | 2.94 X 10$^{-6}$ /8.66 X 10$^{-6}$ /6.402 X 10$^{-6}$ | 1.371 X 10$^{-6}$/ 4.04 X 10$^{-6}$/2.98 X 10$^{-6}$ | 8.14 X 10$^{-10}$/ 8.44 X 10$^{-11}$ /3.69 X 10$^{-9}$ | [12] |
| TaS$_2$ | CW 532nm /671 nm/457 nm | 1.14 X 10$^{-5}$,0.88 X 10$^{-5}$, 0.69 x 10$^{-5}$ cm$^2$/W | ------ | 1.2 X 10$^{-6}$/ 0.9 X 10$^{-6}$/ 0.7 X 10$^{-6}$ | [13] |
| TaSe$_2$ | 532 nm / 671 nm | 8.0 X 10$^{-7}$/3.3 X 10$^{-7}$ (cm$^2$W$^{-1}$) | 1.37 x 10$^{-7}$/1.58 X 10$^{-7}$ | 3.1 X 10$^{-10}$/1.64 X 10$^{-10}$ | [14] |
| GeSe | 532 nm | 4.841 X 10$^{-6}$ (cm$^2$W$^{-1}$) | 2.258 X 10$^{-6}$ | 2.945 X 10$^{-10}$ | [15] |
| Boron NS | CW 457 nm /532nm /671 nm | 1.25 X 10$^{-5}$ / 3.43 X 10$^{-6}$ / 9.45 X 10$^{-6}$ (cm$^2$W$^{-1}$) | 1.75 X 10$^{-7}$ / 0.64 X 10$^{-6}$/ 0.48 X 10$^{-6}$ (esu) | 4 X 10$^{-9}$/1.8 X 10$^{-9}$/ 1.8 X 10$^{-9}$(esu) | [16] |
| SnS NS | CW 532 nm/ 633 nm | 4.531 X 10$^{-5}$/0.323 X 10$^{-5}$(cm$^2$W$^{-1}$) | 2.317 X 10$^{-5}$/0.165 X 10$^{-5}$(esu) | 6.995 X 10$^{-10}$ / 2.037 X 10$^{-10}$(esu) | [17] |
| Bi$_2$S$_3$ | CW 457nm/532 nm/ 671 nm | 3.34 X 10$^{-5}$/1.26 X 10$^{-6}$/1.62 X 10$^{-7}$(cm$^2$W$^{-1}$) | ------ | ------- | [18] |
| MoSe$_2$ | CW 532nm | 3.24 × 10$^{-10}$ W/m$^2$ | -------- | 1.1×10$^{-9}$(e.s.u) | [19] |
| TaAs | 405 nm/ 532 nm/ 671 nm/ 841 nm | ------- | 6.06 × 10$^{-4}$ / 5.68 × 10$^{-4}$ / 5.30 × 10$^{-4}$/ 4.65 × 10$^{-4}$ (e.s.u) | 10.50 × 10$^{-9}$ / 9.86 × 10$^{-9}$ /9.19 × 10$^{-9}$ / 8.07 × 10$^{-9}$ (e.s.u) | [20] |
| **NiTe$_2$(our work)** | CW 405 nm/ 532 nm / 650 nm | 3.22 ×10$^{-5}$/ 6.15 ×10$^{-5}$ / 7.68 × 10$^{-5}$ (W/cm$^2$) | 1.56 × 10$^{-3}$ / 2.99 × 10$^{-3}$ / 3.76 × 10$^{-3}$ | 4.06 × 10$^{-9}$/ 7.79 × 10$^{-9}$/ 9.89× 10$^{-9}$ (e.s.u) | |

Table S2. SSPM formation time of the diffraction pattern as reported in the relevant literature.

| Material | Laser Specification | Solvent | Intensity | Formation time (T) | Ref. |
|---|---|---|---|---|---|
| MoTe$_2$ | 473/532/750 nm (CW) | NMP | 252 W/cm$^2$ | 0.45 s/ 0.6 s/ 0.62 s | [8] |
| MoSe$_2$ | 671 nm (CW) | NMP/ Acetone | 12 W/cm$^2$ | 0.41 s/ 0.22 s | [21] |
| TaAs | 589 nm/ 532 nm/ 473 nm (CW) | NMP | 90 W/cm$^2$ | 2.5 s/ 2.5 s/ 2.3 s | [20] |
| Graphene Oxide | 532 nm (CW) | IPA | ---- | 0.43 s | [22] |
| Black Phosphorus | 700 nm (CW) | NMP | 18.9 W/cm$^2$ | 0.7 s | [11] |
| NiTe$_2$ (This Work) | 405 nm/ 532 nm/ 650 nm | IPA | 2.41 W/cm$^2$, 4.64 W/cm$^2$, 8.18 W/cm$^2$ | 0.122 s, 0.21 s, 0.41 s | |

**Section: S2 Characterization of 2D-hBN**

**XRD:** A distinct peak was identified approximately at 26.7°, corresponding to the (002) plane of the 2D hBN powder. Additionally, a minor peak was observed at 39.05°, which corresponds to the (100) plane. As shown in figure S 3b.

**Raman Spectroscopy:** The exfoliated hexagonal boron nitride(hBN) displays a distinct Raman peak at 1366-1373 cm$^{-1}$ attributed to E$_{2g}$ phonon mode.[23] In this study, the exfoliated 2D hexagonal boron nitride flakes demonstrated a Raman peak at 1367.37 cm$^{-1}$, can be seen in the Figure S 3c.

**AFM:** An atomic force microscope (Agilent Technologies, pico scan 5100, cantilever length of 100 µm) is used to assess the thickness of the exfoliated 2D samples as shown in Figure S 3d. The average thickness was determined to be between 5-10 nm. And the average lateral width is found to be 150-200 nm.

**UV-Visible Spectroscopy:** The Figure S 3i shows the Uv-visible spectroscopy spectra of 2D-hBN. UV-Visible spectroscopy was used to calculate the absorbance of the 2D-hBN. The optical band gap of the 2D-hBN which is found to be 3.24 eV, was calculated using the Tauc plot.

**XPS**: The chemical and compositional analyses of hexagonal boron nitride nanosheets were conducted using X-ray Photoelectron Spectroscopy (XPS). The results distinctly reveal the predominant presence of boron and nitrogen as primary elements, with carbon and oxygen detected as impurity elements in the samples. The chemical analysis of the hBN flakes shows that peaks corresponding to B 1S (Figure S 3j) and N 1S (Figure S 3l), are identified at 190 eV (associated with the B-N bond) and 397.35 eV (associated with the N-H bond) respectively.[24] Further, the XPS spectra of the B 1S is deconvoluted, the presence of supplementary boron-oxygen bonding was detected 191.25 eV as shown in Figure S 3j. Further analysis of the O 1S peak shows evidence of augmentation in boron – oxygen bonding at 532.21 eV[23c], evident in Figure S 3k. This unwanted oxidization might have caused due to sample handling. The aforementioned results are well aligned with the previous reports.

**TEM:** High-resolution transmission electron microscopy was performed at different magnifications to assess the plane orientation. As shown in the Figure S 3e, transparent sheets are seen to be stacked on top of each other, also the grid and the sample are distinguished. An FFT image is determined from one of the zoomed-in images (Figure S 3f), it shows plane direction along the (002) plane, and d spacing is calculated to be 3.45 Å, which corresponds to the (002) plane. Figure S 3g shows an aberration improvised picture made by ImageJ software, a line profile is drawn to record the intensity profile. Green balls indicate nitrogen atoms and blue balls indicate Boron atoms respectively, indicating high-intensity nitrogen atoms than that of boron atoms. A 3D map is shown in Figure S 3h to find the regularity of this intensity variation. Bright-field TEM image of the 2D hBN nanoflakes distinguishing between grid and sample. Inset showing a 3D view of stacking of layers.  Inset showing FFT image of the hexagonal pattern of the (002) plane.  High-resolution core level XPS spectra of B 1S, O 1S, and N 1S.

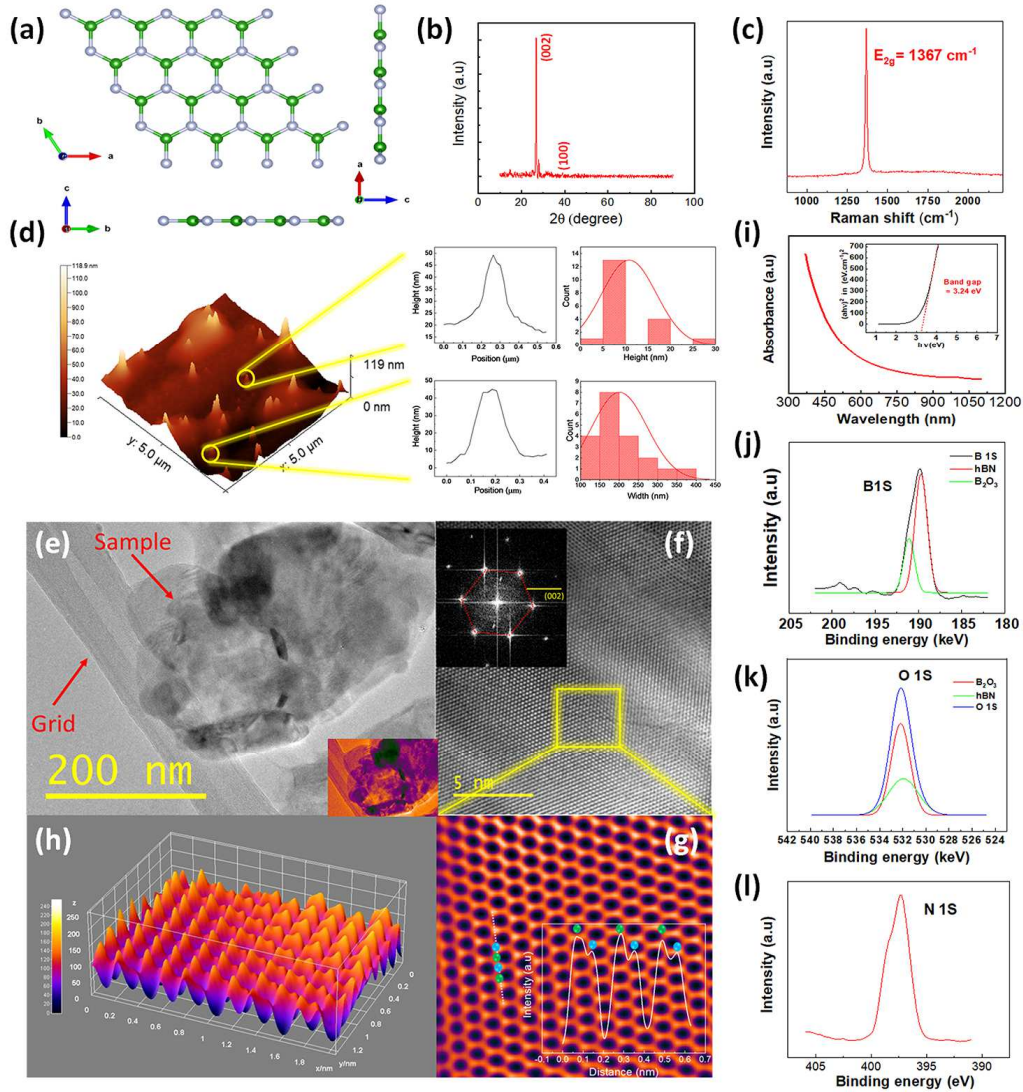

**Figure S3.** a) Atomic structure of 2D h-BN nanosheets. b) XRD pattern of the 2D hBN with sharp (002) peak. c) Raman Spectra hBN with a sharp peak at 1367.37 cm$^{-1}$ corresponding to E$_{2g}$ peak. d) Atomic force microscopic images of 2D NiTe$_2$. Right panel distribution of height profile and lateral dimension of the flakes. e) showing Bright field TEM image of the 2D hBN nanoflakes distinguishing between grid and sample. Inset showing a 3D view of stacking of layers. f) Zoomed in view of the nanoflake Inset showing an FFT image of the hexagonal pattern of the (002) plane. High-resolution core level XPS spectra of j) B 1S, k) O 1S, and l) N 1S.

Table S3: The values of $\chi^{(3)}$, Mobility (μ) & Effective Mass ($m^*$)

| Material & Corresponding Wavelength | $\chi^{(3)}_{monolayer}$ (Third-order nonlinear susceptibility) | Mobility (μ) & Effective Mass ($m^*$) | Reference |
|---|---|---|---|
| Graphene | 1 X 10$^{-3}$ (e.s.u) | [25] | [2, 26] |
| BP | 10$^{-5}$ cm$^2$W$^{-1}$ (e.s.u) | [27] | [11] |
| MoS$_2$ | 1.44 X 10$^{-4}$ (e.s.u) | [28] | [1] |
| WSe$_2$ | 1.371 X 10$^{-6}$/ 4.04 X 10$^{-6}$/2.98 X 10$^{-6}$ (e.s.u) | [29] | [12] |
| TaAs (405 nm/ 532 nm/ 671 nm/ 841 nm) | $6.06 \times 10^{-4}$ / $5.68 \times 10^{-4}$ / $5.30 \times 10^{-4}$/ $4.65 \times 10^{-4}$ (e.s.u)/ $6.06 \times 10^{-4}$ / $5.68 \times 10^{-4}$ / $5.30 \times 10^{-4}$/ $4.65 \times 10^{-4}$ (e.s.u) | [20] | [20] |
| MoSe$_2$ (532 nm) | $1.76 \times 10^{-4}$ (e.s.u) | [30] | [21] |
| WS$_2$ | 8.14 X $10^{-10}$/ 8.44 X $10^{-11}$ /3.69 X $10^{-9}$ (e.s.u) | [31] | [32] |
| MoTe$_2$ | 1.88 X $10^{-9}$ esu 1.3 X $10^{-9}$ esu 1.14 X $10^{-9}$ esu 0.98 X $10^{-9}$ esu (CW 473 nm /532 nm/ 750 nm/ 801 nm) | [33] | [8] |

Table S4: Comparison of the optimized lattice parameters for bulk 1T-NiTe$_2$ using different exchange and correlation functionals.

| xc-functional | a (Å) | c (Å) |
|---|---|---|
| GGA-PBE | 3.895 | 5.254 |
| rVV10 | 4.048 | 5.418 |
| DFT-D3 | 4.021 | 5.289 |
| Obk88-vdW | 4.106 | 5.326 |
| Exp. [34] | 3.859 | 5.275 |

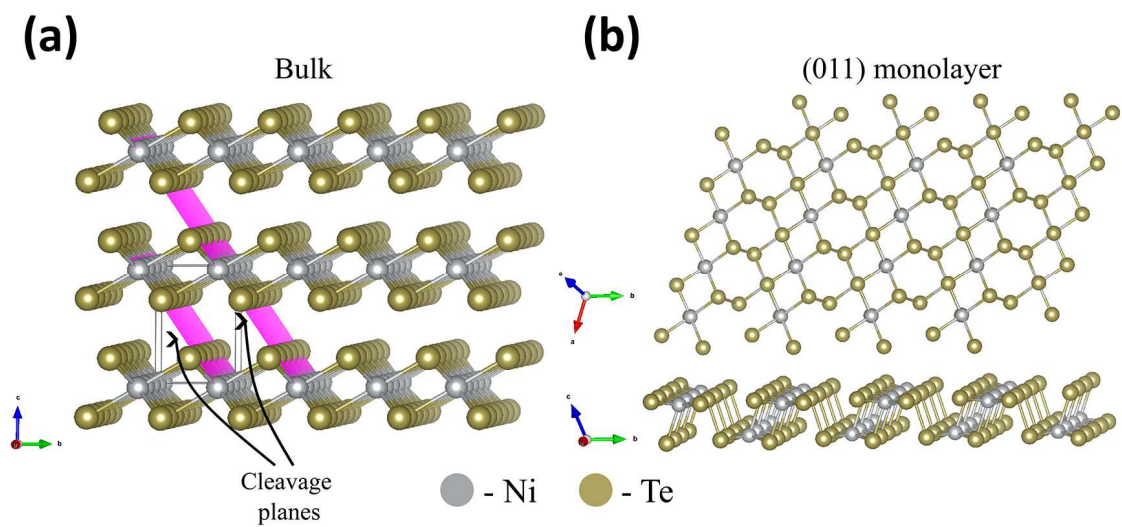

**Figure S4.** a) Bulk supercell with the [011] planes representing the direction of cleavage for constructing the 2D model. b) Top and side view of the optimized 2D (011) monolayer.

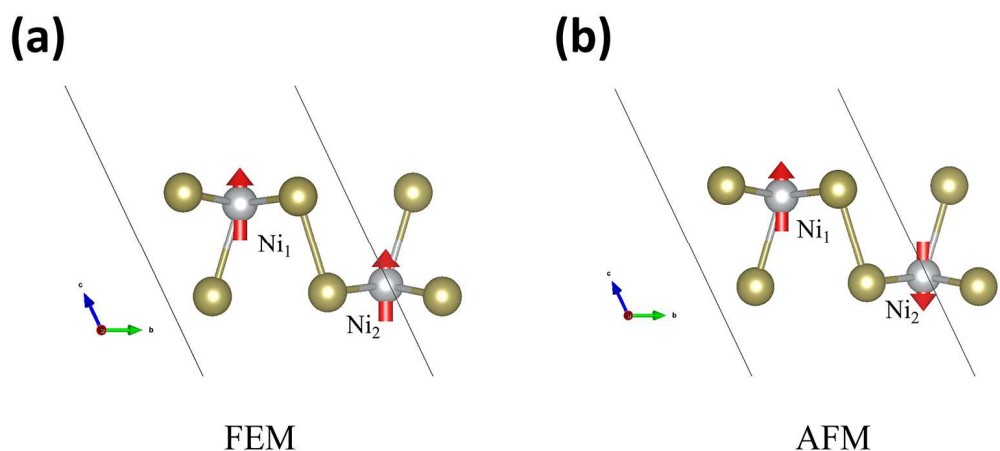

**Figure S5:** a-b) Magnetic configurations considered for optimization of the (011) 2D structure. The red arrows represent the direction of the magnetic moments set on the Ni atoms

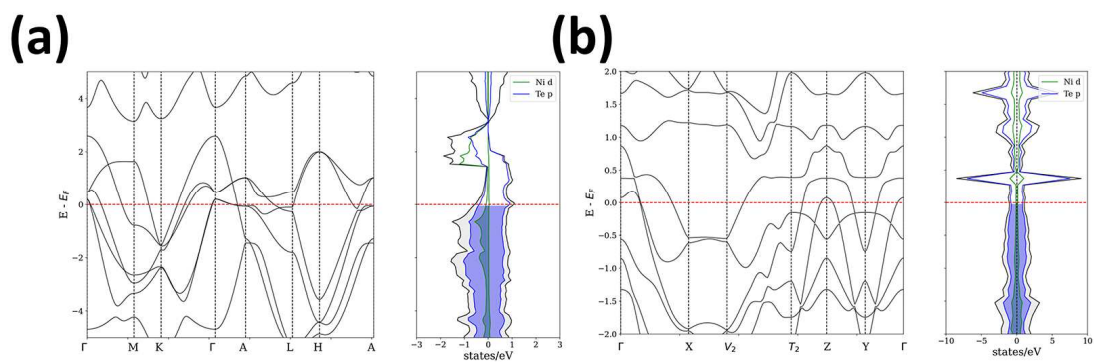

**Figure S6.** Electronic structure of NiTe$_2$. a) Bulk NiTe$_2$ along the conventional path in the BZ and b) electronic band structure for 2D structure.

## Section: S3

Further experimentation with 90° film polarizer is applied to observe whether the dynamic collapse phenomenon is caused by intricate electric field and 2D-NiTe$_2$ coherence or thermally induced convection flow in the fluid. Laser beam with 532 nm is used for the experimentation; after exposing the laser beam with 90° film polarizer the laser beam is focused on the sample. No major difference in evolution time of the diffraction pattern is spotted. Figure S5a is demonstrates the optical setup used to get the diffraction pattern. The noticed change before and after applying polarizer film is noted in the Supporting information Table S5.

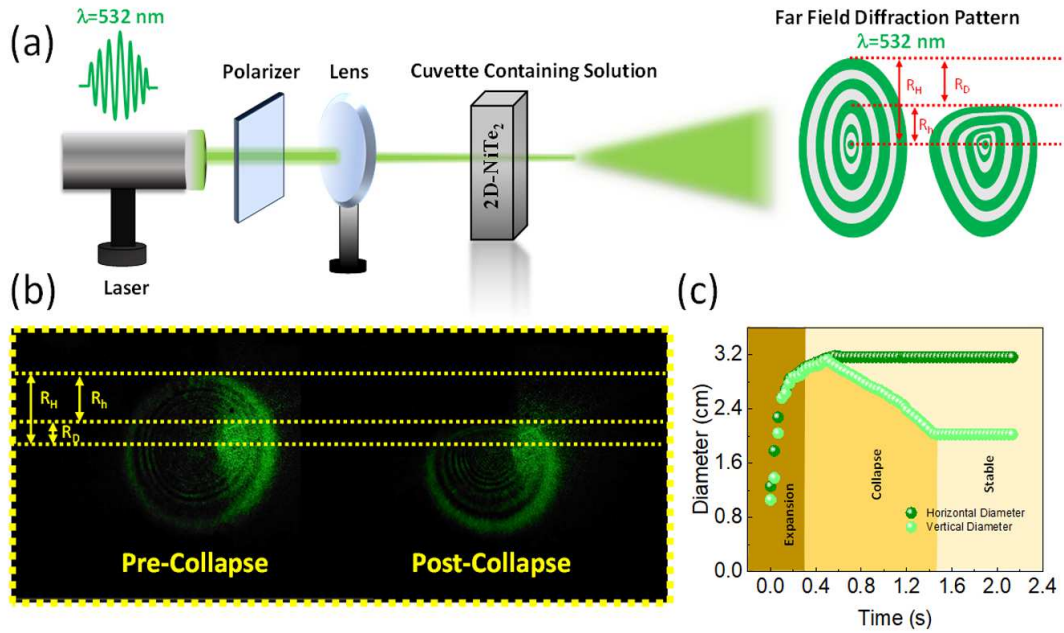

**Figure S5.** Visual representation of the dynamic collapse process. a) Diagrammatic depiction of the collapse phenomenon of the SSPM ring with pre collapse and post collapse picture. c) The evolution of the vertical and horizontal diameter with time for wavelength 532nm.

Table S5: Comparison in dynamic Expansion and Collapse time under different polarization with same wavelength and intensity.

| Laser Specification (532 nm) | Polarization | Time taken for the diffraction pattern to attain maximum Vertical diameter | Time taken for the diffraction pattern to collapse in the Vertical direction |
|---|---|---|---|
| 5.17 W/cm$^2$ | 0 | 0.5 s | 0.833 s |
| 5.80 W/cm$^2$ | 90° | 0.5 s | 0.933 s |

**Section S4:**

Fan et al. developed passive optical diode using two silicon rings of 5 micrometer in radius, although it is a passive device it shows optical non-reciprocal behavior for a wide range of input power level. This compact device showed enhanced performance while backward input power is higher than forward input. Although this device is made of Si based component, it starts to show thermal issue even at low power operation.[35] Feng et al. demonstrated an optical device using two-mode Si waveguide is 200 nm thick and 800 nm wide, this device is capable of showing non-reciprocal light transmission.[36] Wang et al. demonstrated an optical diode which does not require magnetic field or strong input field. The optical device is based on moving structure which relies on a dynamic photonic crystal created inside a three-level electromagnetically produced transparency medium, where the refractive index of a weak probe is altered by the oscillating periodic intensity of a robust standing coupling field with two detuned counterpropagating components.[37] Xu et al. proposed a method to achieve nonreciprocity for a weak input light field using nonlinearity and artificial magnetism. We demonstrate that photons transferred from a linear cavity to a nonlinear cavity (i.e., a linear-nonlinear optical molecule) experience nonreciprocal photon blocking without evident nonreciprocal transmission. The interplay of nonlinearity and synthetic magnetism offers an efficient approach to the development of quantum nonreciprocal devices, magnetism is required to achieve this kind of nonreciprocal light propagation.[38] Floess et al. successfully develop an ultra-thin plasmonic optical rotator in the visible spectrum that facilitates a polarization rotation that is constantly adjustable and switchable by an external magnetic field. The rotator is a hybrid structure that combines a magneto-optical EuSe slab with a one-dimensional plasmonic gold grating. Although magnetism is essential for attaining this kind of nonreciprocal light propagation.[39] Hwang et al. presented a novel hetero-PBG structure with an anisotropic nematic layer interposed between two cholesteric liquid-crystal layers with distinct helical pitches. Authors optically visualized the dispersion relation of this structure, demonstrating the optical diode performance, namely the non-reciprocal transmission of circularly polarized light inside the photonic bandgap areas. The lasing activity achieved during experiment the device was proven to exhibit the diode effect with specific directionality.[40] It is observed that the any magnetism, polarization or any electric field is required to achieve this non-reciprocal light propagation, where in this experimentation the diode is all-optical in nature. Exploiting the reverse saturable absorption property of the 2D-hBN this optical diode is engineered. A 532 nm film polarizer is embedded in the all-optical diode setup to see any change in the output behavior. Here Figure S8a depicts the setup required for the for the above-mentioned experimental procedure. Upper portion reveals the forward bias arrangement and the lower portion of the Figure reveals the reversed bias arrangement for the wavelength λ= 532 nm. Through forward bias arrangement $\frac{dN}{dI}$ is calculated for the wavelength 532 nm and it is found to be 3.35 which is very close to the calculated value of $\frac{dN}{dI}$ of the single cuvette 2D-NiTe$_2$ under the laser illumination of the wavelength 532 nm. The Figure S8 ①-⑩ shows the progression of the laser beam under different intensity.

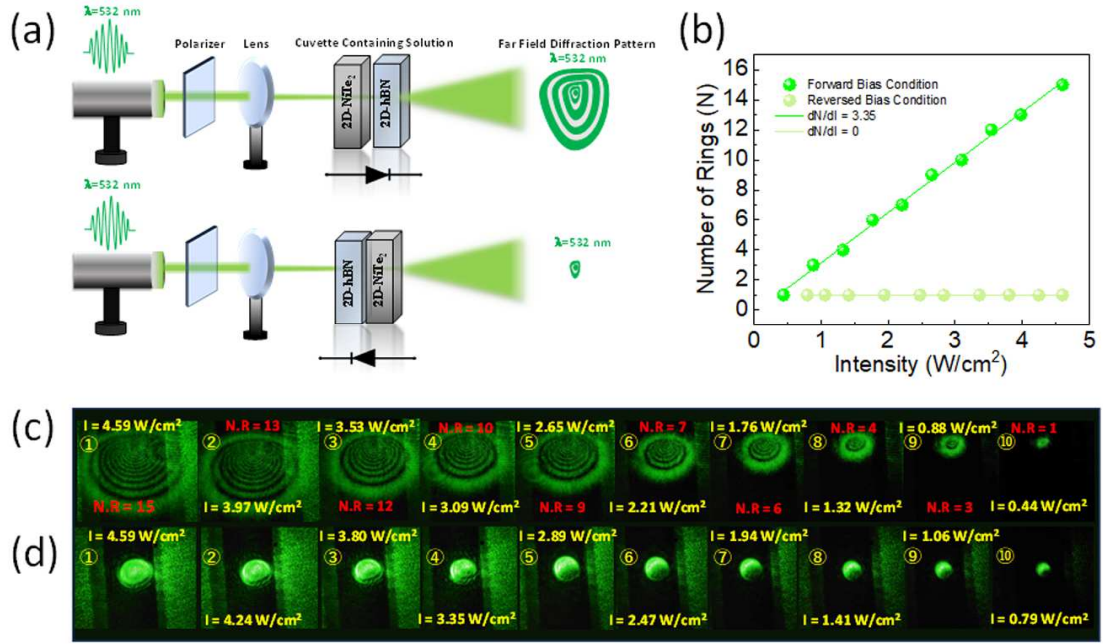

**Figure S8.** a) Figure showing forward and reverse biased analogy in case of the all-optical diode with film polarizer of 532 nm. c) (①-⑩) Shows the diffraction rings vs intensity for forward condition for different wavelengths (532 nm). d) Shows the diffraction rings vs intensity for reversed condition for different wavelengths (532 nm).

**Table S6:** Relative position between the sample and the focal point of the lens in the SSPM experimental setup.

| Wavelength (nm) | Concentration | Cuvette length | $L_2$ (cm) | $L_1$ (cm) | $n_0$ |
|---|---|---|---|---|---|
| 650 | 0.25 | 10 | 10.5 | 9.5 | 1.38 |
| 532 | 0.25 | 10 | 9.9 | 8.9 | 1.384 |
| 405 | 0.25 | 10 | 8 | 7 | 1.39 |
| 650 | 0.13 | 10 | 9.45 | 8.45 | 1.38 |
| 650 | 0.0625 | 10 | 8.4 | 7.4 | 1.38 |
| 650 | 0.25 | 5 | 10 | 9.5 | 1.38 |
| 650 | 0.25 | 1 | 9 | 8.9 | 1.3 |

**Section S5:**

This work is based on the nonlinear optical properties of 2D $NiTe_2$ using SSPM Spectroscopy. The light source used in the SSPM experiment is a CW laser. As per some of the recent work, the SSPM does not originate due to the thermally induced response. The supporting arguments are discussed below.

Statment:1

In the main manuscript in Section 3.6 named Electronic Relation between $\chi^{(3)}_{monolayer}$, Mobility ($\mu$) and Effective Mass ($m^*$), we have discussed the relation between effective mass and carrier's mobility. The carrier's mobility is defined as the carrier's ability to move in the presence of an external field. This is dependent on the effective mass and scattering of the carriers. In the third-order nonlinear optical response case scenario the mobility of the carriers in the coherent laser field establishes the ac nonlocal electronic coherence. Hu et al.[41] expected that $\chi^{(3)}_{monolayer}$ is positively correlated with the carrier mobility. As both of these parameters signify charge storing property (where energy storing mechanism is observed), and $\chi^{(3)}_{monolayer}$ is negatively correlated with the carrier effective mass. The effective mass is a dissipative property (where the energy-releasing mechanism is observed). The light field enhances the motion of the carriers as less scattering was observed during the motion hence higher value of $\chi^{(3)}_{monolayer}$ is observed. This high value of $\chi^{(3)}_{monolayer}$ increases with decrease in wavelength, as the photon energy increases with decreasing wavelength. In the main manuscript, we have summarized the values of $\chi^{(3)}_{monolayer}$ evaluated through SSPM Spectroscopy method vs carrier mobility $\mu$ and effective mass $m^*$. $\chi^{(3)}_{monolayer}$ is an optical parameter, $\mu$ is an electronic property, $m^*$ electronic structure during excited state. This correlation between optical property and electronic property supports laser induced coherence phenomenon.

Statement: 2

Besides the Kerr nonlinearity, variations in the temperature of the medium due to a strong laser beam can alter the refractive index, resulting in similar self-phase modulation as previously reported. Dabby et al.[42] described this phenomenon as the "thermal lens effect". The experiment we have investigated is based on the "Wind Chime Model", which depends on the polarization of suspended nanostructure and subsequent reorientation under the application of a laser beam. Although the thermal lens effect exhibits linear optical response.[43] Then mechanical chopper is implemented to verify the above-mentioned claim.

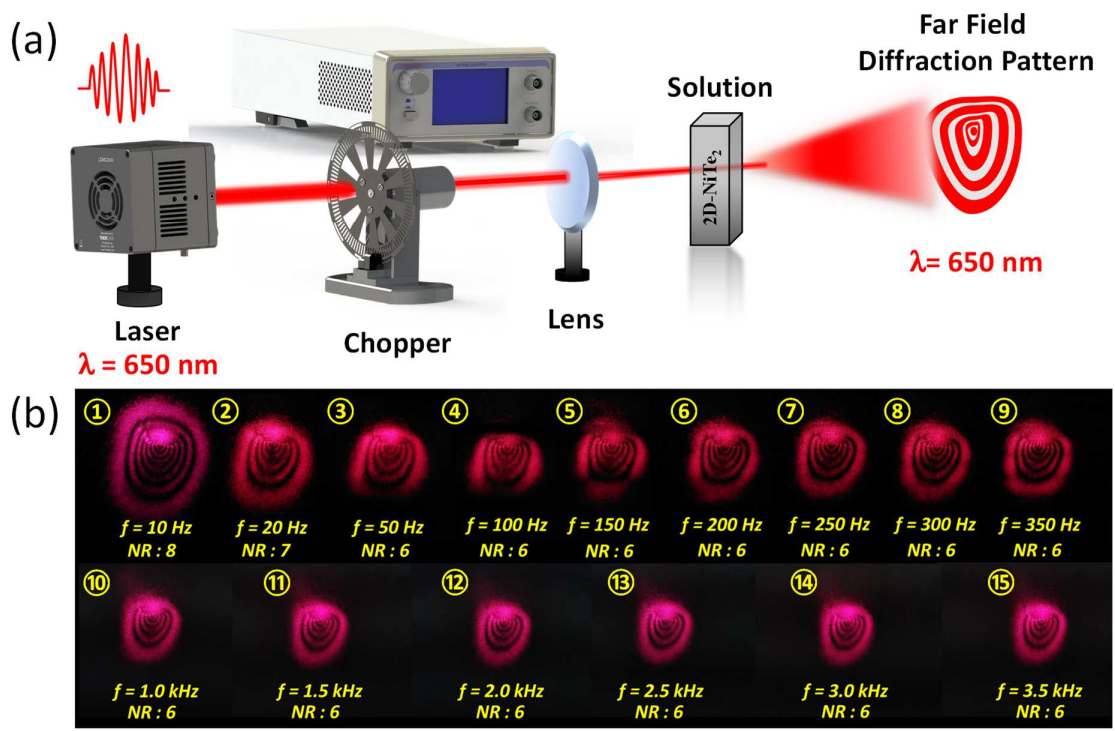

**Figure S9.** (a) Schematic representation of the SSPM setup with mechanical chopper operating between (0-3.5 kHz). (b) SSPM pattern at different mechanical chopper modulation frequencies (10-350 Hz and 1.0-3.5 kHz)

Figure S9a shows the optical setup incorporating a mechanical chopper. The chopper is initially operated at frequency range between 10-350 Hz at intensity 6.05 W/cm$^2$. For the same intensity the chopper is operated at higher frequency 1.0-3.5 kHz to observe the change in the number of rings, no substantial change was found. Hence, we can deduce that the electronic coherence phenomenon which gives rise to NLO response dominates the thermal lens effect.

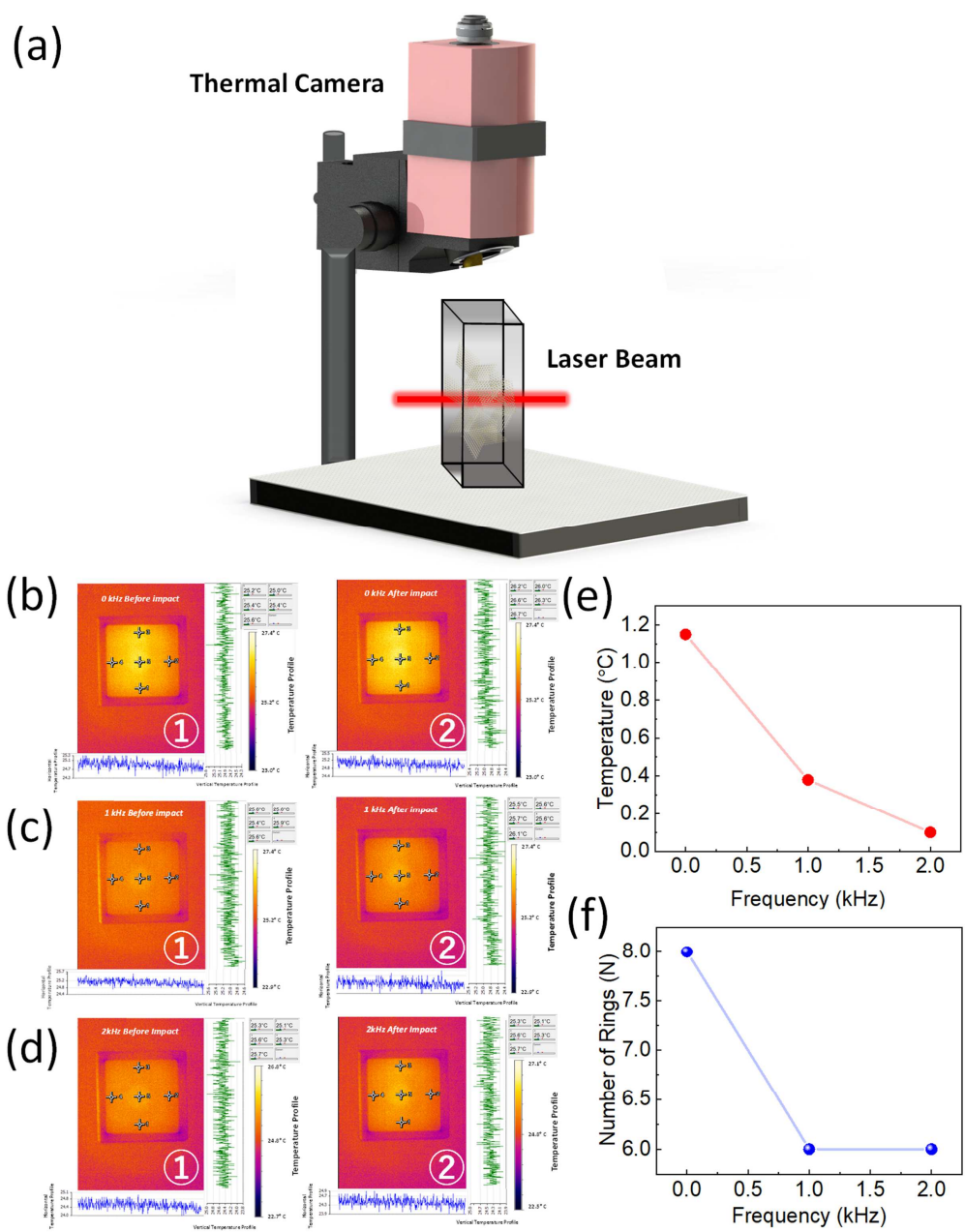

**Figure S10.** (a) Thermal camera setup for measurement of the upper view of the thermal profile of the cuvette. (b)- (c) -(d) Thermal profile of the cuvette before and after the laser impact for frequencies 0 kHz, 1 kHz, and 2 kHz. (e) The temperature difference before and after laser impact. (f) Number of Rings vs frequency of the mechanical chopper.

For the same intensity, the thermal camera was applied in the following setup depicted in Figure S10a. For different chopping frequencies the thermal interaction of the laser with the solvent is measured, where the "Thermal lens effect" is predicted to arise. For different frequencies 0 kHz, 1 kHz, and 2 kHz the temperature difference is measured before laser impact and after laser impact. Figure S10b①, c①, and d① depict the upper view of the cuvette before laser impact for frequencies 0 kHz, 1kHz, and 2 kHz. Figures S10b②, c②, and d② depict the upper view of the cuvette after laser impact for frequencies 0 kHz, 1kHz, and 2 kHz. In the Figure S10e, the temperature difference is measured before laser impact and after laser impact with chopper frequency. It was observed that as the chopping frequency increases the laser interaction decreases with the solvent and the localized heating of the solvent is reduced. For that temperature decrease is observed with an increase in chopping frequency. It was observed that after 2 kHz (also noticed) temperature increase is very low or nonexistent. Yet the number of rings stays the same even after the increase in chopping frequency, shown in Figure S9b ⑫-⑮. The initial decrease in the number of rings is caused to the decrease in intensity of the laser beam, hence less electronic coherence and a lesser number of rings generated. Although the temperature is decreased the ring number stayed the same as seen in Figure S10f, leading to the conclusion that the SSPM phenomenon is not a thermally induced phenomenon.

The above statements provide in-depth justification that SSPM is not a thermally induced phenomenon, caused by the "thermal lens effect".

**Section S6:**

Here authors have discussed the rationale behind not using 1000 nm laser. In the study authors have taken measurements at the wavelengths of 650, 532, and 405 nm. The third order nonlinear susceptibility can be achieved at the wavelength 1000 nm. Although the mechanism of the Spatial self phase modulation depends on the polarization of the 2D-nanostructure upon light-matter interaction. Previously Huang et al.[20] measured the value of $\chi^{(3)}_{total}$ and $\chi^{(3)}_{monolayer}$ for TaAs and found to be lower compared to 405, 532, and 671 nm. The value of $\chi^{(3)}_{total}$ and $\chi^{(3)}_{monolayer}$ are found to be increasing with decreasing wavelength. As the wavelength increases energy per photon decreases following the equation $E = h\frac{C}{\lambda}$. Here h is Planck's constant, C is the speed of light, and λ is wavelength of the incoming light. As the shorter wavelength laser beam has more energy per photon the polarization capability is considered to be greater. The same is observed in this study also, the data is presented in the Table S7.

Table S7: The value of $n_2$, $\chi^{(3)}_{total}$, $\chi^{(3)}_{monolayer}$ calculated using SSPM Spectroscopy in recent literature

| Wavelength (nm) | Concentration (mg mL$^{-1}$) | L (mm) | dN/dI (cm$^2$ W$^{-1}$) | $n_2$ (cm$^2$ W$^{-1}$) | $\chi^{(3)}_{total}$ (e.s.u) | $\chi^{(3)}_{monolayer}$ (e.s.u) |
|---|---|---|---|---|---|---|
| 650 | 0.25 | 10 | 1.36 | $3.22 \times 10^{-5}$ | $1.56 \times 10^{-3}$ | $4.06 \times 10^{-9}$ |
| 532 | 0.25 | 10 | 3.20 | $6.15 \times 10^{-5}$ | $2.99 \times 10^{-3}$ | $7.79 \times 10^{-9}$ |
| 405 | 0.25 | 10 | 5.27 | $7.68 \times 10^{-5}$ | $3.76 \times 10^{-3}$ | $9.8 \times 10^{-9}$ |

For same level of polarization of the nanostructure at a specific concentration higher wavelength laser beam requires higher intensity. For this reason, high threshold value of laser intensity observed (i.e. the minimum laser intensity required to generate first diffraction ring) to be high for higher wavelengths. Huang et al.[20] used 841 nm for SSPM experiment and encountered error due to high threshold value. Experimentation with 1000 nm laser will introduce a great amount of threshold, which would lead to incorrect evaluation of the parameter $\frac{dN}{dI}$. This incorrect value of $\frac{dN}{dI}$ may affect the determination of the value $n_2$, $\chi^{(3)}_{total}$, and $\chi^{(3)}_{monolayer}$. Also, using higher value of wavelength results in low[8,1] value of $\frac{dN}{dI}$, for realization of photonic diode robust light matter interaction is expected. Hence the choice of wavelength is restricted to three wavelength (λ= 650, 532, and 405 nm). Using lower value compared to 405 nm might cause enhanced vertical distortion, which can affect the calculation of $\frac{dN}{dI}$. Also, at lower wavelength the reverse bias format will stop functioning as the laser beam will be able polarize the 2D hBN flakes, and interaction will cause diffraction pattern to emerge.

**Section S7:**

Low number of rings has been obtained for 1 mm thick cuvette. This might have happened due to decreased light matter interaction. Further verification using 2 mm cuvette might have been the correct choice. For 2 mm thickness cuvette at the intensity 9.8 W/cm$^2$ the number of rings in the diffracted pattern is found to be 7, which is found to be higher than that of 1 mm. For 1 mm cuvette thickness at intensity 9.5 W/cm$^2$ the number of rings is found to be 5. Hence a decrease in the number of rings is observed as cuvette length decreases. Figure S 11a depicts the change in the number of rings in the diffraction pattern with intensity. The corresponding $dN/dI$ is calculated to be 0.72, which is very close to the $dN/dI$ for 1 mm thickness cuvette. Figure S 11f shows the calculated value of $dN/dI$ for the cuvette thickness of 10, 5, 2, and 1 mm at constant concentration (0.25 mg mL$^{-1}$) and wavelength (650 nm). The corresponding value of $dN/dI$ is found to be 1.36, 1.16, 0.72, and 0.71 cm$^2$W$^{-1}$.

The threshold value is defined as the minimum incoming laser intensity required to produce the first ring of the SSPM Phenomenon. As with previous research, we establish a threshold value for each experimental instance for calculating $dN/dI$. The line has to traverse the origin (0,0) to get an exact slope value. Thus, the slope of the line associated with the threshold value should be almost identical to the optimal slope needed to get the $\chi^{(3)}_{total}$ value. The third-order nonlinear susceptibility is exactly proportional to the gradient of the ring number concerning the laser intensity (i.e. $\lim_{I \to \infty} N/I$). This section presents a null value at (0,0) and computes the linear slope, thereafter comparing it with the realistic scenario. But for the case of cuvette with the thickness 1 mm is found to have different $dN/dI$ when the threshold value is considered. Figure S 11b and Figure S 11c shows the value of $dN/dI$, while considering the threshold value and not taking it. A discrepancy in the value is observed, hence dual gradient fit method is adopted.[44] The lowest gradient fit slope corresponding to the threshold value is found to be 0.21 and 0.56 corresponding to the 1 mm and 2 mm cuvette, which shows low threshold value for 1 mm cuvette. This low slope corresponds to high threshold value. Corresponding slopes are depicted in the Figure S 11d and Figure S 11e.

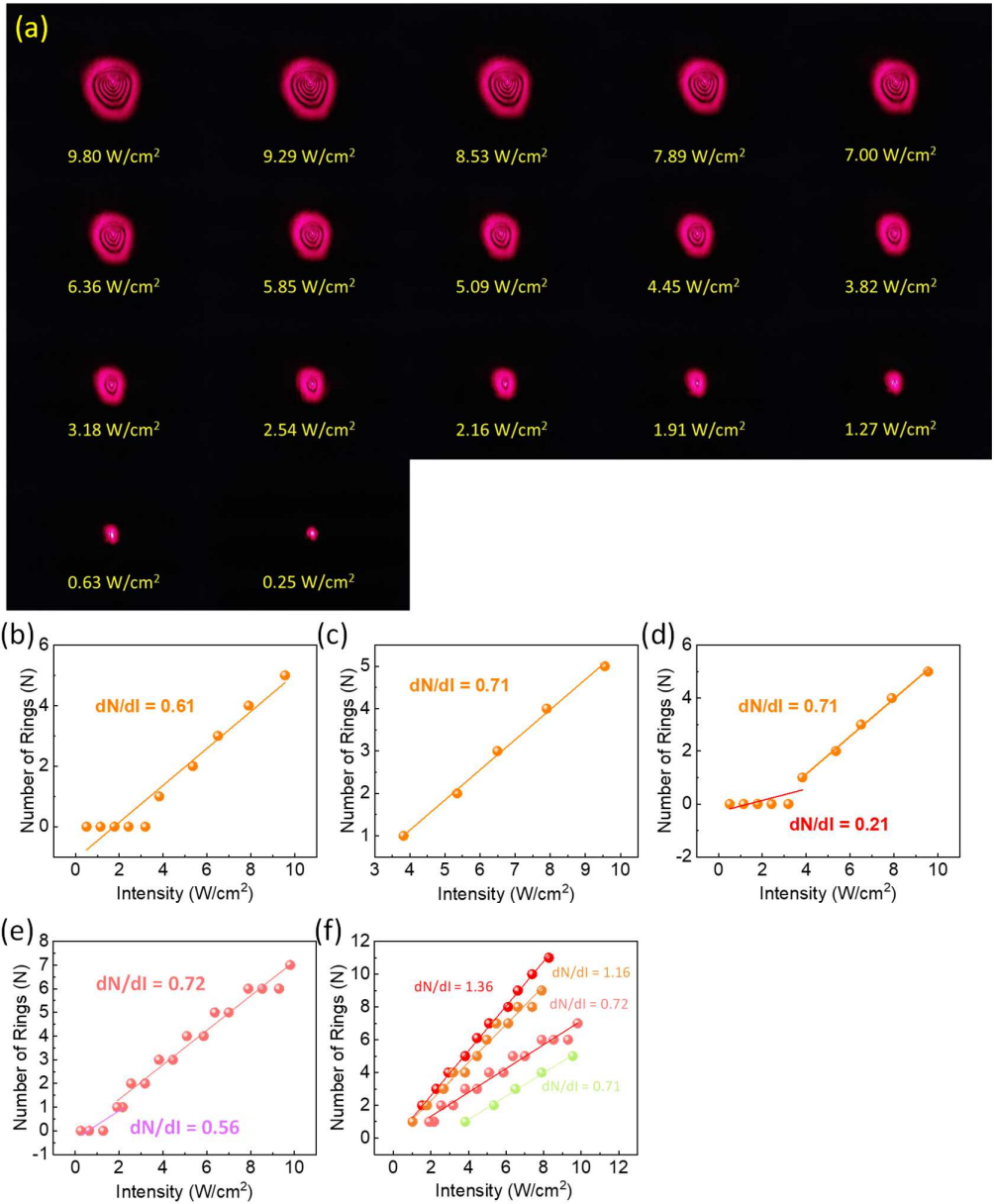

**Figure S11:** Figure depicts the dN/dI for the cuvette length 1 mm b) with taking consideration of (0,0) point and threshold value c) without it. d) dN/dI value d) dN/dI value of the 1 mm cuvette system considering dual gradient fit method. e) dN/dI value of the 2 mm cuvette system considering dual gradient fit method. f) The variation of the number of diffraction rings vs intensity for different cuvette lengths (10 mm, 5 mm, 2 mm, and 1 mm).

**Section S8:**

In this Section Authors have elaborately discussed the selection interest of particular wavelengths 650, 532, and 405 nm for All-Optical diode application.

1. Here authors have discussed the rationale behind not selecting the IR range wavelength for SSPM Spectroscopy. As discussed in the Comment 1 the nonlinear optical response of the 2D-NiTe₂ at 1000 nm is less compared to other wavelength. For realization of nonlinear optical diode a strong light matter interaction is expected at a particular wavelength. At a particular wavelength if the incoming intensity of the laser beam is high the quantified response as the value of $\frac{dN}{dI}$ is expected to be also high. As discussed in Comment 1, following the previous literature the response for 1000 nm is expected to low. Any wavelength after 650 nm is expected to show high threshold value. An enhanced value of $\frac{dN}{dI}$ suggests that the proposed nonlinear photonic diode can exhibit a more pronounced effect of nonreciprocal light propagation. For In this scenario, the incoming laser beams (405, 532, and 650 nm) must possess sufficient intensity to interact with 2D NiTe₂, thereby causing the optical Kerr effect and subsequently generating diffraction rings as a result of the narrow bandgap. Low value of $\frac{dN}{dI}$ would suggest non optimal reciprocal light propagation. Also, at lower wavelength the reverse bias format will stop functioning as the laser beam will be able polarize the 2D hBN flakes, and interaction will cause diffraction pattern to emerge.

2. Similar Comparison Method is included to further add advantage for selection of the particular wavelengths (650, 532, and 405 nm). Recent reports show use of wavelengths 671, 532, 457, and 405 nm for Similar comparison method. Similar comparison method can be used to calculate the value of $n_2$ for NiTe₂ based photonic diode and other similar semiconducting nanostructure, where nonreciprocal light propagation has been achieved. The nonlinear refractive index can be defined as,[45]

$$n_2 = \frac{\lambda}{2n_0 L_{eff}} \cdot \frac{dN}{dI}$$

Here $\lambda/2n_0 L_{eff}$ is a constant, $\lambda$, $n_0$, $L_{eff}$ are the waelength of the incoming laser beam, linear refractive index, and effective optical path length of the incoming laser beam inside the cuvette. To estimate the nonlinear refractive index of a 2D NiTe₂-based photonic diode, it is necessary to conduct a comparative analysis between the 2D material in question and other materials that have a well-established nonlinear refractive index.

Similar contrast (S) is defined as,

$$S = 1 - D = \frac{|n_{21} - n_{22}|}{n_{21}}$$

$$S = 1 - \frac{\left|\frac{\lambda}{2n_0 L_{eff}} \frac{N_1}{I_1} - \frac{\lambda}{2n_0 L_{eff}} \frac{N_2}{I_2}\right|}{\frac{\lambda}{2n_0 L_{eff}} \frac{N_1}{I_1}} = 1 - \frac{\left|\frac{N_1}{I_1} - \frac{N_2}{I_2}\right|}{\frac{N_1}{I_1}}$$

Here D is the difference contrast, $n_{21}$ and $n_{22}$ represent the nonlinear refractive index of the heterostructure obtained from the foward bias and reverse bias conditions respetively. Here similar contrast is measured with other 2D materials, and the information of the 2D material coefficients are documented in Table S8.

Table S8: The value of material, nonlinear index, and Similar contrast calculated using SSPM Spectroscopy.

| Material | Nonlinear Refractive Index ($cm^2 W^{-1}$) | Similar Contrast (%) | Reference |
| --- | --- | --- | --- |
| $MoS_2$ | $\approx 10^{-7}$ | 74% | [32] |
| $Bi_2Se_3$ | $\approx 10^{-9}$ | 58% | [10] |
| SnS | $\approx 10^{-5}$ | 90% | [17] |
| Sb | $\approx 10^{-6}$ | 85% | [5] |
| Graphdiyne | $\approx 10^{-5}$ | 92% | [46] |
| CuPc | $\approx 10^{-6}$ | 75% | [21] |
| Graphene | $\approx 10^{-5}$ | 95% | [21] |
| $SnS_2$ | $\approx 10^{-9}$ | 54% | [46] |
| $NiTe_2$ | $\approx 10^{-5}$ | 91.8% | |

From the Figure S12 it can be concluded that the 2D NiTe$_2$ exhibits a nonlinear refractive index comparable to Sb and SnS, indicating a $n_2$ range of $10^{-5}$ cm$^2$ W$^{-1}$. Which was earlier confirmed through experimentation through SSPM spectroscopy. This proves validity and repeatability of the SSPM experiment. The 2D-NiTe$_2$ exhibits a high comparable contrast ($\approx$90%) compared to SnS, indicating a significant nonlinear refractive index ($\approx 10^{-5}$ cm$^2$ W$^{-1}$).

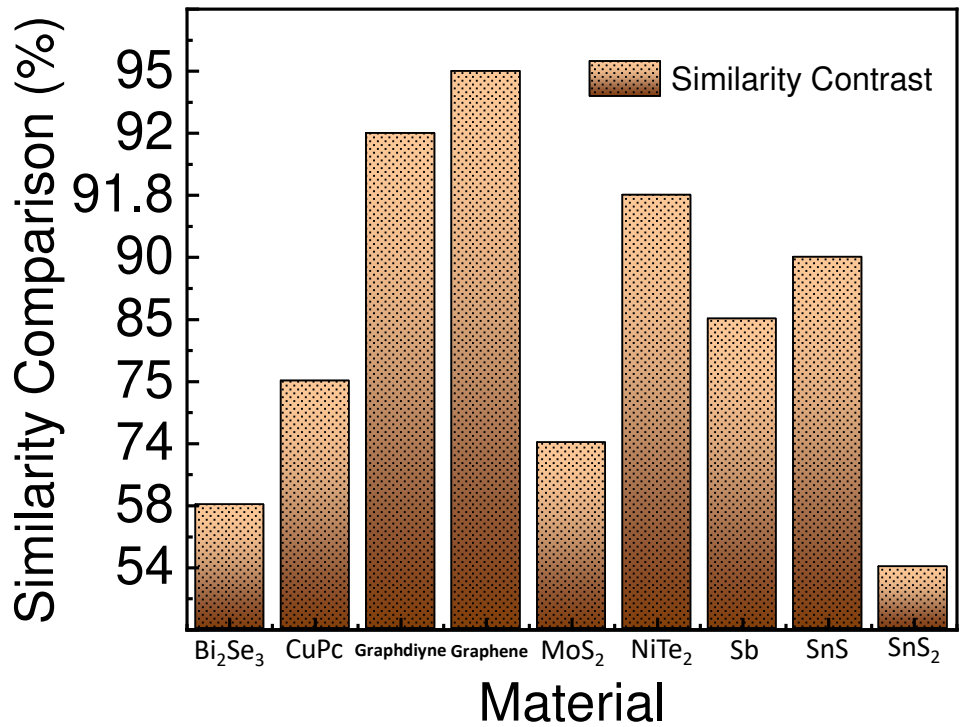

Figure S12: Similarity comparison method for $n_2$ estimation in all photonic diode

**Section S9:**

Surface plasmon polaritons (SPPs) are linked electromagnetic and electronic collective excitations localized at a metal-dielectric interface.[47] Surface plasmon polaritons exhibit distinctive electromagnetic phenomena, including diminished wavelength and localized field amplification in relation to incident light. These characteristics, including optical tunability and control within specified spectral bandwidths, facilitated extensive technological applications.[48] While traditional metals are the standard method for generating surface plasmon polaritons (SPPs) [47], unconventional conductors such as transparent oxides[49], high-temperature superconductors[50], strongly correlated oxides[51], and graphene[52] have recently demonstrated the capability to host SPPs with customized properties due to their heightened sensitivity to doping, temperature, and electric and magnetic fields Also, an enhanced effective nonlinear optical response may be attained via plasmonic phenomena. Such effects result from coherent oscillations of conduction electrons at the surface of noble metal complexes.[53] Plasmonic excitations may enhance nonlinear optical phenomena via many mechanisms. The interaction of light to surface plasmons may produce intense localized electromagnetic fields.[53-

[54] This concept is relevant to improving the nonlinearity of the metal itself (intrinsic response) or that of an adjacent material (extrinsic response). A surface plasmon polariton (SPP) is a travelling surface wave at the continuous metal-dielectric interface related to the reflection, transmission, or absorption of light. The electromagnetic field diminishes rapidly on either side of the contact, facilitating subwavelength confinement next to the metal surface. The number of optically sensitive topological materials has recently been augmented by the identification of Weyl semimetals (WSMs) and Dirac semimetals (DSMs).[55] WSMs and DSMs permit fermions that follow to the Weyl/Dirac quantum field equation at specific electronic band crossing locations (designated as Weyl/Dirac nodes), so they can be considered a three-dimensional analogue of graphene. For Type II DSM the Dirac cone is excessively tilted in the k direction. In the type-I class, the Dirac cone is not excessively inclined.[56] The $NiTe_2$ nanostructure being semi-metallic (bulk) in nature it possesses Dirac node, for which the conductivity of the of the bulk nanostructure can be between graphene and metals. Rizza et al. have distinctive electromagnetic response of $NiTe_2$ in the near-infrared and optical spectrum, attributed to its pronounced anisotropy resulting from the existence of Lorentz-violating Dirac fermions.[57] Utilizing electron energy loss spectroscopy and ab initio density functional theory, authors detected several ENZ surface plasmons, occurring around photonic topological transitions (about 0.79, 1.64, and 2.22 eV), whose existence corroborates the typical topological photonic characteristics of $NiTe_2$.

A concentrated laser beam directed at the semi-metallic surface of the $NiTe_2$ nanostructure may initiate evanescent plasmonic waves that propagate to both ends of the nanostructure, and the waves reflected by the ends will generate standing-wave patterns. This standing wave may interact with the electric field of the incoming laser beam and follow wind chime model, which could give rise to SSPM phenomenon.

In the context of 3-dimensional emerging type of Dirac fermions inside topological Dirac/Weyl semimetals, the existence of a linear Dirac conic band structure is anticipated to provide optical characteristics akin to those of graphene. This discovery promptly indicated a promising possibility to replicate the nonlinear plasmonics of graphene in three-dimensional topological materials, free from the constraints of graphene's two-dimensionality. Zhang et al. derived the nonlinear optical conductivities of three-dimensional massless Dirac fermions inside the quantum mechanical interband domain. Results indicated that 3D topological Dirac/Weyl semimetals concurrently preserve the three-dimensional structural benefits of traditional bulk metals and exhibit pronounced optical nonlinearity characteristic of Dirac quasiparticles.[58] This straightforward yet effective model encapsulates the fundamental physics of the three-dimensional gapless Dirac cone in Dirac and Weyl semimetals. Authors have computed the real component of the optical conductivity through Kubo linear response theory.[59]

To demonstrate its optical responsiveness, optical susceptibilities is defined as[58],

$$\chi^{(n)}(\omega) = i\sigma^{(n)}/n\epsilon_0\omega$$

Due to high conductivity of the charge carriers a high value $\chi^{(n)}$ is noted. of The factor to the polarization can be expressed as,

$$P(t) = \epsilon_0[\chi^{(1)}E + \chi^{(2)}E^2 + \chi^{(3)}E^3 \dots]$$

Here the material NiTe$_2$ is centrosymmetric material, the even susceptibility terms are nullified $\chi^{(2)}$, $\chi^{(4)}$, $\chi^{(6)}$, and so on. The value of $\chi$ increases on the increase on polarizability, hence enhanced light matter interaction is observed.